\documentclass[papersize,journal]{IEEEtran}
\usepackage{amsmath,amsfonts}
\usepackage{algorithmic}
\usepackage{algorithm}
\usepackage{array}
\usepackage[caption=false,font=normalsize,labelfont=sf,textfont=sf]{subfig}
\usepackage{textcomp}
\usepackage{stfloats}
\usepackage{url}
\usepackage{verbatim}
\usepackage{graphicx}
\usepackage[colorlinks=true,linkcolor=black,anchorcolor=black,citecolor=black,filecolor=black,menucolor=black,runcolor=black,urlcolor=black]{hyperref}
\usepackage{color}
\usepackage{scalefnt}
\usepackage{cite}
\begin{document} 
\title{\huge{On the SER Performance of ZF and MMSE Receivers in Pilot-Aided Simultaneous Communication and Localization} }

\author{Shuaishuai Han,~\IEEEmembership{Student Member,~IEEE,} Emad Alsusa, 
\IEEEmembership{Senior Member,~IEEE,} Arafat Al-Dweik, \IEEEmembership{Senior Member,~IEEE}

\thanks
{Shuaishuai Han and E. Alsusa are with the Department of Electrical and Electronic Engineering,
University of Manchester, Manchester M13 9PL, U.K. (e-mail: {shuaishuai.han@outlook.com,}
e.alsusa@manchester.ac.uk). Arafat Al-Dweik is with the 6G Research Centre, Department of Computer
and Information Engineering, Khalifa University of Science and Technology, Abu Dhabi, United Arab Emirates. (e-mail: {, arafat.dweik@ku.ac.ae).}}\thanks}

\date{}
\maketitle

\begin{abstract}
\textbf{In this paper, a symbol error rate (SER) analysis is provided to evaluate the impact of localization inaccuracy on the communication performance under Zero-Forcing (ZF) and Minimum Mean-Square Error (MMSE) equalizers. Specifically, we adopt a pilot-aided simultaneous communication and localization (PASCAL) system, in which multiple drones actively transmit signals towards the base station (BS). Upon receiving the signal, the BS estimates the drones' location parameters to reconstruct the channel matrix, which is then utilized for ZF and MMSE equalization. As the channel matrix is characterized by the estimated parameters associated with the target’s location and the matrix inversion involved in ZF and MMSE further complicates the analysis, obtaining a closed-form SER expression becomes intractable. Thus, a tightly approximated SER expression is respectively derived for ZF and MMSE by using a hybrid approximation method incorporating Neumann approximation and Taylor approximation. Our analysis reveals several important design insights: first, the average SER of drone $k$ for both ZF and MMSE can be affected by the localization errors from all drones including drone $k$; second, the average SER of ZF is unaffected by the estimation inaccuracy of range, whereas the average SER of MMSE is influenced by it; third, ZF and MMSE is the most susceptible to the influence of angle estimation errors compared to the other localization errors; fourth, ZF is highly sensitive to localization errors and may be even worse than maximal ratio combining (MRC) under some conditions of significant estimation errors. Numerical simulation results verify our findings and also validate the accuracy of the analysis across a wide range of system parameters.} 
\end{abstract}
\begin{IEEEkeywords}
Symbol error rate (SER), Zero-Forcing (ZF), Minimum Mean-Square Error (MMSE), pilot-aided simultaneous communication
and localization (PASCAL)
\end{IEEEkeywords}

\section{Introduction}
Many promising technologies have emerged alongside ongoing research efforts toward next-generation wireless networks, such as ultra-massive MIMO and integrated sensing and communication (ISAC) \cite{ISAC_liu, ISAC_liu2, ISAC_Mu, ISACJarrah1}. Compared to separated sensing and communication systems, ISAC enables communication and sensing functionalities to share both spectrum, hardware platforms and signal processing modules, thus enhancing resource utilization efficiency \cite{ISAC_Lyu1, ISAC_Xu}. Nevertheless, most of current ISAC studies ignore the fact that the targets to be localised may have their own transceivers. In fact, these targets can actively transmit signals to the base station (BS) by using their transceivers, thereby enabling not only reliable communication but also localization through pilot symbols embedded in the signal. Thus, the pilot-aided simultaneous communication and localization (PASCAL) system is introduced in \cite{ISAC2_Han}. 

ISAC is gaining momentum owing to its promising potential for future development; nevertheless, certain aspects in ISAC still warrant further improvement. ISAC schemes typically employ separate signals for communication and sensing. For instance, in \cite{ISAC_Qi,  ISAC_ouyang2}, the ISAC BS transmits sensing signals toward the targets and then extracts sensing information from the returning echoes. Meanwhile, it also receives uplink signals from users to support communication functionality. Nevertheless, the simultaneous reception of communication and sensing signals may introduce a mutual interference issue. In addition, a weight factor is employed to adjust the sensing and communication performance in \cite{ISAC_Qi}. The system resources are allocated to sensing and communication in proportion to the weight factor. In the PASCAL system, as shown in Fig. ., the signals transmitted from the targets are employed not only for acquiring the targets' location, but also to facilitate communication with them. By sharing the same signal for both functions, the potential interference between communication and sensing can be avoided. Furthermore, it is unnecessary to allocate different portions of resources to communication and sensing in PASCAL, the signal jointly used by communication and sensing can fully utlize all available resources. Thus, the utilization efficiency of resources can be improved in PASCAL compared to the ISAC works in \cite{ISAC_Qi,  ISAC_ouyang2}. Attributable to the aforementioned advantage, PASCAL holds significant potential to replace ISAC in applications especially with limited resources, such as drone-based communication \cite{drone_qiu}.  reference

\subsection{Literature Review}

In the field of joint sensing and communication, extensive research efforts have been devoted to enhance both communication and sensing performance of the system in the past several years. Among the numerous research, one fundamental and important topic is communication-sensing trade-off issue. For instance, under a fixed resource budget, allocating more resources to sensing will reduce the communication performance. As a consequence, many trade-off analyses are conducted to maximise the overall performance of the ISAC systems \cite{ISAC_Tian, ISAC_Xiong}. In \cite{ISAC_Tian}, a trade-off problem between Cramér-Rao bound (CRB) and sum transmission rate, which are respectively employed to measure the localization and communication performance, is investigated. This work try to minimize CRB with the communication rate constraint. In \cite{ISAC_Xiong}, both the maximum achievable data rate with a minimum CRB constraint and the minimum achievable CRB under a communication-rate constraint are explored. In addition, it is been revealed in \cite{ISAC_Xiong} that the trade-off in ISAC systems is determined not only by resource allocation, but also depends on data modulation schemes. In \cite{ISAC_Zou}, a trade-off problem to minimize the transmit power, while satisfying both sensing and communication requirements, is solved in the fluid Antenna Systems. In \cite{ISAC_Fodor}, the trade-off analyses between angle of arrival (AOA) and symbol estimation are conducted in both deterministic model and stochastic model. The deterministic model treats the source waveforms as non-random, while the stochastic model assumes that the waveforms follow zero-mean Gaussian distribution. An insight obtained in \cite{ISAC_Fodor} is that the ISAC trade-off is more intricate in the stochastic model, whereas the deterministic waveforms allow a nearly independent allocation of communication and sensing resources. Other trade-off research is provided in \cite{ISAC_an, ISAC_hua}. The trade-off research in \cite{ISAC_Tian, ISAC_Xiong, ISAC_Zou, ISAC_Fodor, ISAC_an, ISAC_hua} is beneficial to provide insights in designing ISAC systems. Nevertheless, the interaction mechanism between sensing and communication has been overlooked in many literature including \cite{ISAC_Tian, ISAC_Xiong, ISAC_Zou, ISAC_Fodor, ISAC_an, ISAC_hua}. In fact, the localization results can be employed to reconstruct the channel by using the parametric channel estimation method \cite{para_Zia} since localization and communication functions jointly employ the same channel in the PASCAL system. Thus, it is essential to investigate the mechanism from which different localization parameters affect the communication performance \color{black}. 

Some performance evaluation are conducted based on equalizers like maximum ratio combining (MRC), zero forcing (ZF), or minimum mean square error (MMSE) in the ISAC systems. For instance, in \cite{ISAC_Parihar}, the analytical results for outage probability, system throughput and ergodic sum rate are provided after applying MRC to process the received signals to analyze a ISAC system with active multiple-functional reconfigurable intelligent surfaces (RISs). In \cite{ISAC_Liu}, a closed-form upper bound for bit error rate (BER) and outage probability are respectively derived based on a ZF equalizer to evaluate the performance of uplink NOMA-ISAC system.  In \cite{ISAC_Nguyen}, a achievable rate is derived in a closed form expression based on MRC and ZF schemes for massive MIMO ISAC systems. The analyses in \cite{ISAC_Parihar, ISAC_Liu, ISAC_Nguyen} are beneficial to evaluate the advantages of their respective proposed systems. In these analyses, the performance with various equalization techniques is affected by channel estimation errors. Nevertheless, the impact of channel estimation inaccuracy on equalizer performance has already been extensively studied.


\subsection{Motivation and Contributions}

Although PASCAL brings significant advantages over the ISAC designs in \cite{ISAC_Qi, ISAC_ouyang2}, the mechanisms through which localization inaccuracy affects communication performance remain insufficiently explored. This is also a notable gap in the current ISAC-related research, which focues more on the trade-off between sensing and communication functions, while neglects the bidirectional interactions or synergistic benefits between sensing and communication. In our previous work in \cite{ISAC2_Han}, the SER anaysis is provided based on a MRC equalizer. Even if  prior work on \cite{ISAC2_Han} has successfully revealed the influence of localization inaccuracy on MRC, the mechanism of MRC is simplistic (i,e., multiply the received signal with the conjugate of the estimated channel) and may not fully capture the actual performance of a communication system under various scenarios. In specific, MRC maximizes the output signal-to-noise ratio (SNR) but it does not take interference into account.  Other equalizers like ZF and MMSE are more essential and more frequently used than MRC in the applications with severe user interference or in scenarios that requires high performance.  Thus, the aim of the paper is to provide a SER analysis of the PASCAL system under the ZF and MMSE receivers. More importantly, we aim to investigate how the localization imperfections (i.e., angle, range and Doppler frequency estimation errors) affect the performance of ZF and MMSE. The previous work in \cite{ISAC2_Han} indicates that only angle and Doppler frequency estimation defects impact the performance of MRC. In addition, the average SER for drone $k$ is only affected by the estimated location errors from drone $k$. In this paper, we would like to investigate whether the conclusion remains valid for ZF and MMSE. The contributions of the paper are listed as follows:
\begin{enumerate}
\item The SER analyses are provided to quantify the influence of localization defects on the performance of ZF and MMSE, respectively. In the PASCAL system, the channel matrix is composed of estimated location parameters, each following independent Gaussian distributions, and is estimated using an alternating optimization maximam likelihood (AO-ML) algorithm. Afterwards, the estimated channel matrix is fed into the ZF and MMSE equalizers for SER analyses. 

\item Due to the significant correlation issue caused by estimated location parameters and the complex inverse calculation involved in ZF and MMSE, a hybrid approximation method consisting of a $N$th-order Neumann
approximation and a $R$th-order Taylor approximation is employed to derive the average SER. Since ZF is affected by only angle and Doppler frequency estimation errors, whereas MMSE is influenced not only by these two factors but also by range estimation errors, the derivation of MMSE is more complex than that of ZF. 

\item The mechanism in which localization errors affects the performance of communication performance with ZF and MMSE is investigated. In specific, we notice that the average SER for drone $k$ of ZF and MMSE are affected by the localization defects from all drones. In addition, compared to other localization errors, ZF and MMSE 
are most significantly influenced by the angle estimation errors. The above-mentioned influence mechanism is verified by our SER analsis. These findings are helpful to provide some insights in designing communication systems that highly rely on location accuracy, for instance, location-aware service.

\item Simulation results are provided to demonstrate the influence of localization inaccuracies on ZF and MMSE. The communication performance with ZF and MMSE improves with the reduction of localization inaccuracies. In addition, ZF is highly susceptible to localization inaccuracies, in some scenarios, may perform worse than MRC. Subsequently, the high consistency between the simulated and analytical results indicate the accuracy of the our analysis. 
\end{enumerate}

$Notations$: ${[ \cdot ]^*}$, ${[ \cdot ]^T}$ and ${[ \cdot ]^H}$ refers to the complex conjugate, transposition and Hermitian transposition.  $\left \| \cdot \right \|_2$ indicates the 2-norm. $\mathbb{E}[\cdot]$ represents the statistical expectation. $\hat{\boldsymbol{ \psi}}$ denotes a estimation vector of unknown parameters. 

\begin{figure}[ptb]
	\centering
	\includegraphics [scale=0.43]{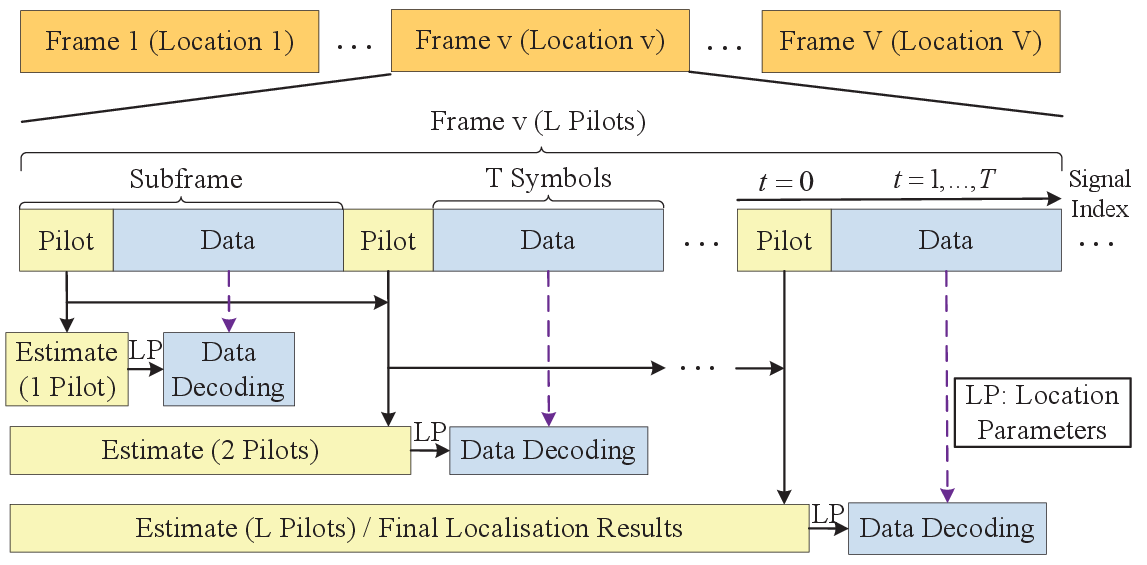}
\caption{Frame structure of the PASCAL system.}%
	\label{PASCAL}

\end{figure}

\section{System Model}
\label{system model}
In PASCAL, we consider a multiuser single-input multiple-output (MU-SIMO) system, where a number of $K$ moving single-antenna drones are employed to actively transmit signals to a BS with $N$ antennas. Once the BS receives the signals, it can extract the location parameter (i.e., direction of arrival (DOA), range and Doppler frequency) from the received signals with the assistance of pilot signals to complete the localization of the drones. Meanwhile, the estimated location parameters are employed to reconstruct the channel by using the parametric channel estimation method, which is then fed into various equalizers to complete data decoding. The parametric method models the wireless channel through a set of underlying physical parameters—such as angles and path gains, as opposed to the non-parametric approach, which considers the channel response as a whole. Given that the location parameters has already been estimated during the localization process, we can directly employ them to obtain the channel state information (CSI). Thus, the parametric channel estimation method is will-suited for the PASCAL system.


The PASCAL can be applied to a variety of scenarios. One scenario is GPS-denied environments \cite{GPS_denied} for instance sub-bridge areas or scenarios subjected to jamming, in which the GPS information may not be available. Another one is high-accuracy localization scenario since commercial GPS systems may be insufficient to meet operational and safety requirements. In above-mentioned scenarios, the devices equipped with transceiver, such as drones, can actively can actively transmit signals to the BS for localization and rely on a number of pilot signals to guarantee the localization accuracy. In the meantime, the signals from the drones can be employed for communication with the BS. Thus, an efficient resource (e.g., power) utilization can be achieved by using the same signal for both functions. In addition, since the system experiences one-way path loss, rather than the two-way path loss suffered by the localization method based on echoes, the localization performance can be improved. In the paper, the ground-to-air (G2A) model is employed. Due to the the mobility and flexibility of drones, the Line-of-Sight (LoS) probability is high in the G2A model, thus the LoS assumption is employed in the ISAC literature like \cite{ISAC_Lyu}. In addition, the Non-LoS (nLoS) paths may lead to a phenomenon commonly referred to as ghost targets in radar images \cite{ghost}. As a consequence, the LoS/NLoS identification technique, such as the one in \cite{LoS Extract}, is employed to identify and extract the LoS component prior to performing localization.


\subsection{Communication-Localization Model}

To support efficient communication and localization in the PASCAL system, the frame structure shown in Fig. \ref{PASCAL} is employed. In Fig. \ref{PASCAL}, each frame is composed of $L$ subframe, in which each subframe contains one pilot and $T$ symbols. In the $l$ subframe, $l$ pilots are employed to estimate the location parameters of the drones, which are then utilized for data decoding in that subframe. We assume that the movement of the drones within each frame is negligible given that the frame length is very short. For instance, a frame comprising 100 symbols occupies only 0.1 ms at a symbol rate of 
$1\times 10^6$ symbols/s. In contrast, the minimum coherence time of a LoS channel for a mobile drone traveling at 180 km/h and operating at 2 GHz is approximately 3 ms. Thus, the assumption is practical and also consistent with that of \cite{localization_gao}. In Fig. \ref{PASCAL}, the frame length is flexible and thus the frame structure can be better accommodate the dynamic positional variations of drones. In addition, the number of symbols allocated for data and pilot within each subframe can be flexibly configured, enabling different pilot densities to accommodate various application requirements. For example, in high-mobility scenarios, only 4–6 data symbols may be placed between successive pilot symbols to ensure accurate localization and channel acquisition.    

The PASCAL employs the same signal to complete the both communication and localization functions. The $t$th received signal vector within the $l$th subframe can be written as 
\begin{equation} 
 \setcounter{equation}{1} 
	\label{received signal matrix2}
	{\bf{  y}}_{t,l} = {\bf{A}}{{\boldsymbol{\omega}}(l)}  {\bf{s}}_{t,l}+ {\bf{  n}}, 
\end{equation}
where ${\bf{s}}_{t,l}\in \mathbb{C}{^{K \times 1}}$ refers to the transmitted signal vector for $t\in\{0,...,T\}$ and $l\in\{1,...,L\}$.  In particular, ${\bf{s}}_{0,l}$ represents the pilot signal vector corresponding to $t=0$ for localization purpose, whereas ${\bf{s}}_{t,l}$ for $t \geq 1 $ indicates the data-bearing signal vector.  ${\bf{s}}_{t,l}$ can be given by ${\bf{  s}}_{t,l} = [\sqrt{{{P}}_{1}}{s_{t,l,1}},...,\sqrt{{{P}}_{K}}{s_{t,l,K}}]^T$, where ${{{P}}_{k}}$ indicates the transmit power of drone $k$ and $s_{t,l,k}$ represents the $t$th transmit signal within $l$th subframe corresponding to drone $k$. ${\bf{s}}_{0,l}$ can be denoted by ${{\bf{s}}_{0,l}}=[ \sqrt{{{P}}_{1}},...,\sqrt{{{P}}_{K}}  ]^T$ since ${{{s}}_{0,l,k}}=1$ for $k\in\{1,...,k\}$, which is a frequently used normalisation method by the literature such as \cite{assumption1}. ${\bf{  y}}_{t,l}\in \mathbb{C}{^{N \times 1}}$ denotes the signal vector received at the $N$-antenna BS. ${\bf{  n}} \in \mathbb{C}{^{N \times 1}}$ refers to a vector consisting of additive white Gaussian noise (AWGN). 

${\bf{A}}{{\boldsymbol{\omega}}(l)}\in \mathbb{C}{^{N \times K}}$ refers to the LoS channel response. ${\boldsymbol{\omega}}(l)$ denotes a diagonal matrix consisting of the effect of path loss and Doppler frequency, which can be given by 
\begin{equation}
   {{\boldsymbol{ \omega}}(l)} \!\! =\! \!  \mathrm{diag}\bigg\{\! \eta_1{\exp\! \bigg\{ \! {j2\pi \frac{ f_{D\! , 1}}{f_s}}}\! \bigg\},\! ...,\! \eta_K{
\exp\! \bigg\{\! j2\pi \frac{ f_{D\! , K}}{f_s}\! \bigg\}}\!  \bigg\}, 
\end{equation}
where ${\eta _k}$ denotes the free space path loss, which can be given by ${\eta _k}=\frac{\lambda}{4 \pi d_k}$. $d_k$ refers to the distance from drone $k$ to the BS and $\lambda$ indicates the wavelength. $f_{D,k}$ denotes the Doppler frequency of drone $k$, which is directly associated with its velocity. ${\bf{A}}$ represents the array manifold of the BS, which can be given by ${\bf{A}} =\big[{{\bf{a}}}({\theta _1}) ,...,{{\bf{ a}}}({\theta _K}) ]$. In this paper, a BS composed of a uniform linear array (ULA) with an inter-element spacing of half a wavelength is employed, thus, the steering vector ${{\bf{a}}}({\theta _k})$ can be given by
\begin{equation} 
	\begin{array}{l}
		  {{\bf{a}}}({\theta _k}) = [{{{a}}}_{1}({\theta _k}),...,{{{a}}}_{N}({\theta _k})]^T,
	\end{array}
\end{equation}
where $		\displaystyle{{{a}}}_{n}({\theta _k})\overset{\triangle}{=} {\exp\{- j2\pi  ({n}-1) d_0\sin {\theta _k}/\lambda}\}\;\forall n\in\{1,..,N\}$, in which $\theta_k$ denotes the DOA of drone $k$.  

\section{Alternating-Optimization ML Estimator for localization}
\label{AO-ML}
As shown in Fig. \ref{PASCAL}, $l$ pilot signals (i.e., ${\bf{s}}_{0,1}$, ..., ${\bf{s}}_{0,l}$ ) can be employed to perform localization in the $l$th subframe, and thus the pilot signal vector can be given by 
\color{black} 
\begin{equation}
{{\bf{ y}}_1} = {\left\{\Big\{{{\bf{A}}}{\boldsymbol{\omega}}(1){\bf{s}}_{0,1}\Big\}^T,...,\Big\{{{\bf{A}}}{\boldsymbol{\omega}}({l}){\bf{s}}_{0,l}\Big\}^T\right\}^T} + {{\bf{ n}}_1}, 
\end{equation}
where ${{\bf{ n}}_1}$ indicates the corresponding AWGN vector. The PDF of ${{\bf{ y}}_1}$ in the presence of AWGN noise can be given by 
\begin{equation}
   f({{\bf{ y}}_1}|\boldsymbol{\psi} ) = {{{{\pi ^{-N{l}}}\det ({\bf{\Gamma }})}}}^{-1}{{\rm{e}}^{ - {{\left[{{\bf{ y}}_1} - \boldsymbol{\mu} \right]}^H{{\bf{\Gamma }}^{-1}}\left[{{\bf{ y}}_1} - \boldsymbol{\mu} \right]}}}, \color{black}
\end{equation}
where $\boldsymbol{\psi} = {[{{\boldsymbol{\theta }}^T},{{\boldsymbol{d }}^T},{{\boldsymbol{f_{D}}}^T}]^T}$. ${\boldsymbol{\theta }}$ and ${\boldsymbol{d }}$ refer to vectors composed of angle and range parameters of $K$ drones, while ${\boldsymbol{f_{D}}}$ denotes a vector consisting of Doppler frequency parameters of the drones. $\boldsymbol{\mu}$ and $\bf{\Gamma }$ indicate the mean vector and covariance matrix of ${{\bf{ y}}_1}$, respectively.  $\boldsymbol{\mu}={\left\{\{{{\bf{A}}}{\boldsymbol{\omega}}(1){\bf{s}}_{0,1}\}^T,...,\{{{\bf{A}}}{\boldsymbol{\omega}}({l}){\bf{s}}_{0,l}\}^T\right\}^T}$.
Based on the PDF of ${{\bf{ y}}_1}$, the maximum likelihood estimation (MLE) can be denoted by 
\begin{align}
[ \hat {\boldsymbol{\psi}} ] &= \mathop {\arg \max }\limits_{{\boldsymbol{\psi}} } \ln f({\bf{ y}}_1|{\boldsymbol{\psi }}) \nonumber
= \mathop {\arg \min }\limits_{{\boldsymbol{\psi}}} ||{\bf{ y}}_1 - {\boldsymbol{\mu }}||_2^2,
\end{align}
where $ {\boldsymbol{\psi}}={[{{{ {\boldsymbol \theta}^T }}},{{{ {\boldsymbol d}^T }}},{{{ {\boldsymbol f}^T_{D}}}}]^T}$ contains the location parameters corresponding to all drones.    

Due to the computational burden of MLE, the alternating optimization technique is employed. The joint estimation of a number of parameters can be divided into multiple subproblems, each solved by optimizing one group of variables while keeping the remaining variables fixed. Without loss of generality, we consider estimating one drone's location parameters in each individual problem so that we have 
\begin{align}
\label{AOML}
[\hat  {\boldsymbol{\psi}}_k ] &= \mathop {\arg \min }\limits_{ {\boldsymbol{\psi}}_k} ||{\bf{ y}}_1 - {\boldsymbol{\mu }}(\hat {\boldsymbol{\psi}}_i)||_2^2,
\end{align}
where ${\boldsymbol{\psi}}_k= {[{{{ { \theta}_k }}},{{{ { d}_k }}},{{{ { f}_{D,k}}}}]^T}$ is composed of the location parameters of drone $k$, whereas ${\boldsymbol{\psi}}_i= {[{{{ { \theta}_i }}},{{{ { d}_i }}},{{{ { f}_{D,i}}}}]^T}$ contains the other drones' location parameters for $i \ne k$. The information of ${\boldsymbol{\psi}}_i$ in the subproblem can be obtained from their initial values or the estimated values from last iteration. Thus, ${\boldsymbol{\psi}}_i$ is known in \eqref{AOML}, which is denoted by ${ \hat {\boldsymbol{ \psi}}}_i$.  The alternating iterations over all subproblems until convergence yield the final estimates of the parameters corresponding to all the drones. It is worth mentioning that each subproblem can be solved by using exhaustive search or gradient descent method. To guarantee the convergence speed of the alternating optimization algorithm, the initialization method in \cite{initialization} can be employed. 

In \cite[Sec. IV]{ISAC2_Han}, it has been found that the estimation errors for location parameters follow independent Gaussian distributions. As a consequence, the PDF of the estimation errors for the location parameter corresponding to the $k$th drone can be given by 
\begin{equation}
    f(\Delta {\boldsymbol{  \psi}}_k) = \frac{1}{{\sqrt {2\pi } \sigma_\psi }}{{\rm{e}}^{ - \frac{1}{2}{{\left(\frac{\Delta {\boldsymbol{  \psi}}_k}{\sigma_\psi }\right)}^2}}},
\end{equation}
where $\Delta {\boldsymbol{  \psi}}_k\in\{\Delta { \theta _K}, \Delta { d_{k}}, \Delta { f_{D,k}} \}$ and ${\sigma_\psi }$ refer to the standard deviation of the distribution, which equals to RMSE of the estimated location parameter $\Delta {\boldsymbol{  \psi}}_k$ as 
\begin{equation}
    \sigma_\psi=\text{RMSE}\overset{\triangle}{=} \sqrt{{\mathbb{E}}[ (\hat \psi_{k}-\psi_k)]^2}.
\end{equation}

The Gaussian PDF of the estimation errors can be proved in \cite[Theorem 7.1]{estimation theory}, which indicates that the estimated parameter by MLE asymptotically follows a Gaussian distribution with a mean equal to its actual value if the log-likelihood function is differentiable and the Fisher information is non-zero. In addition, the independence between different estimation errors can be proved in \cite[Theorem 7.3]{estimation theory}. 

\section{Data Detection}
In this paper, we focus on providing SER analysis to evaluate the effect of localization errors on data decoding in the PASCAL system and investigating how localization defects influences different types of equalizers.

\subsection{Data Detection with Maximum Ratio Combining (MRC) Equalizer}
When employing a Maximum Ratio Combining (MRC) equalizer, the received signal vector ${{\bf{ y}}_{t,l}}$ should be multiplied by ${{\bf{\hat W}}_\mathrm{MRC}}$ as ${\bf{ x}}^{\mathrm{MRC}}_{t,l} = {{\bf{\hat W}}_\mathrm{MRC}}{{\bf{ y}}_{t,l}}$, in which ${{\bf{\hat W}}_\mathrm{MRC}}={\bf{\hat H}}^H$. ${\bf{\hat H}}=\hat {\bf{A}}{{\boldsymbol{\hat \omega}}(l)}\in \mathbb{C}{^{N \times K}}$ refers to the channel matrix composed of estimated location parameters obtained from the localization algorithm in Sec. \ref{AO-ML}, where $\bf{\hat A }$ and ${{\boldsymbol{\hat \omega}}(l)} $ denote the vectors composed of the estimated location parameters.  $\bf{\hat A }$ can be given by ${\bf{ \hat A}} = \big[{{\bf{ a}}}({\hat \theta _1}) ,...,{{\bf{  a}}}({\hat \theta _K}) ]$, while ${{\boldsymbol{\hat \omega}}(l)} $ is written as 
\begin{equation}
   \!  {{\boldsymbol{\hat \omega}}(l)} \!\! =\! \!  \mathrm{diag}\bigg\{\! \eta_1\! \big(\hat d_1\big){\exp\! \bigg\{ \! {j2\pi \frac{\hat f_{D\! , 1}}{f_s}}}\! \bigg\},\! ...,\! \eta_K\! \big(\hat d_K\big){
\exp\! \bigg\{\! j2\pi \frac{\hat f_{D\! , K}}{f_s}\! \bigg\}}\!\!  \bigg\},
\end{equation}
where ${\hat \theta _k}$, $\hat d_k$ and ${\hat f_{D,k}} $ refer to the estimated location parameters, which can be expressed in forms of localization errors as ${\hat \theta _K}={ \theta _k}+\Delta { \theta _k}$, $\hat d_k=d_k+\Delta { d_{k}}$ and ${\hat f_{D,k}}={ f_{D,k}}+\Delta { f_{D,k}}$. 

\subsection{Data Detection with Zero-Forcing Equalizer}
\label{localization and communication}

 By using Zero-Forcing (ZF) equalizer, the received signal ${{\bf{ y}}_{t,l}}$ should be multiplied by ${{\bf{\hat W}}_\mathrm{ZF}}$ as ${\bf{ x}}^{\mathrm{ZF}}_{t,l} = {{\bf{\hat W}}_\mathrm{ZF}}{{\bf{ y}}_{t,l}}$, in which ${{\bf{\hat W}}_\mathrm{ZF}}$ represents the Moore–Penrose inverse (i.e., pseudoinverse) of ${\bf{\hat H}}$ and thus ${{\bf{\hat W}}_\mathrm{ZF}}={\bf{\hat H}}^{\dagger}=({\bf{\hat H}}^{H}{\bf{\hat H}})^{-1}{\bf{\hat H}}^{H}$. 

Due to the complexity of the inverse calculation for ${\bf{\hat H}}^{H}{\bf{\hat H}}$ in ${{\bf{\hat W}}_\mathrm{ZF}}$, especially in cases where ${\bf{\hat H}}$ is a function of multiple random variables, the Neumann expansion is invoked to simplify the computational complexity. By denoting ${\bf{G}}_{\mathrm{ZF}}={{\bf{\hat H}}^{H}{\bf{\hat H}}}$, ${\bf{G}}_{\mathrm{ZF}}$ can be approximated by using the $R$th Neumann series as
\begin{align}
\label{Neumann series ZF}
    {\bf{G}}^{-1}_{\mathrm{ZF}}&\approx\sum\limits_{{{r}} = 0}^{R}  (-{\bf{G}}^{-1}_{d,\mathrm{ZF}}{\bf{G}}_{e,\mathrm{ZF}})^{r}{\bf{G}}^{-1}_{d,\mathrm{ZF}},
\end{align}
where ${\bf{G}}_{d,\mathrm{ZF}}=N\bf{I}$ and ${\bf{G}}_{e,\mathrm{ZF}}$ denote the matrices composed of diagonal elements and off-diagonal elements of ${\bf{G}}_{\mathrm{ZF}}$, respectively. It should be mentioned that the approximation errors can be ignored when $R$ is sufficiently large.

By utilising the Neumann series in \eqref{Neumann series ZF}, ${{\bf{\hat W}}_\mathrm{ZF}}$ can also be written as 
\begin{align}
\label{Neumann series ZF21}
    {{\bf{\hat W}}_\mathrm{ZF}}\approx\sum\limits_{{{r}} = 0}^{R}  (-{\bf{G}}^{-1}_{d,\mathrm{ZF}}{\bf{G}}_{\mathrm{ZF}}+{\bf{I}})^{r}{\bf{G}}^{-1}_{d,\mathrm{ZF}}{\bf{\hat H}}^{H}.
\end{align}

By invoking the binomial theorem, ${\bf{\hat W}}_\mathrm{ZF}$ can be simplified to 
\begin{align}
    {{\bf{\hat W}}_\mathrm{ZF}}=(-1)^r\sum\limits_{{{r}} = 0}^{R}\sum\limits_{{{\tilde k}} = 0}^{r}{\left(\! {\begin{array}{*{20}{c}}
\!\!{{r}}\!\!\!\\
\!\!{{\tilde k}}\!\!\!
\end{array}} \!\right)}\frac{(-{\bf{I}})^{r-\tilde k}{\bf{G}}^{\tilde k}_\mathrm{ZF}}{(N)^{\tilde k+1}}{\bf{\hat H}}^{H},
\end{align}

\begin{figure*}[!b]
\hrulefill
 \setcounter{equation}{29} 
  \vspace{-0.2cm}
 \begin{equation}
\begin{array}{*{20}{l}}
 \label{nu}
\nu_\mathcal{D}\!\!=\!\! (-1)^r\sum\limits_{ p = 1}^K \sum\limits_{{{r}} = 0}^{R}\sum\limits_{{{\tilde k}} = 0}^{r}{\left( {\begin{array}{*{20}{c}}
\!\!{{r}}\!\!\!\\
\!\!{{\tilde k}}\!\!\!
\end{array}} \right)}\sum\limits_{{{v}} = 0}^{\tilde k}{\left( {\begin{array}{*{20}{c}}
\!\!{{\tilde k}}\!\!\!\\
\!\!{{v}}\!\!\!
\end{array}} \right)} \frac{(-1)^{r-\tilde k}\alpha^{\tilde k-v}}{(C_\mathcal{D})^{\tilde k+1}} \!\!\!\! \sum\limits_{ i_1,...,i_{v+1}\in S_1}\sum\limits_{ q_1,...,q_{v}\in S_2}\prod\limits_{a=1}^{v}\eta_k(\Delta d_k)\eta^2_{q_a}(\Delta d_{q_a}) \eta_{p}\sqrt{P_p} \vspace{0.3cm}\\  \ \ \   
\times\exp\{\smash{  j\color{blue}2\pi\color{black}\big\{\underbrace{ \scriptstyle \frac{ d_0}{\lambda}[{(i_1-1) \sin ({\theta _k} + \Delta {\theta _k}) - (i_{v+1}-1)\sin {\theta _{p}}+(i_{a+1}-i_a)\sin {(\theta _{q_a}+\Delta \theta _{q_a})}}]}_{\Theta_2}+\frac{m_{p}-1}{M}+\frac{f_{D,{p}}-f_{D,k}-\Delta f_{D,k}}{f_s}\big\}}\}
 \vspace{-0.4cm}
\end{array}  
 \end{equation} 
  \vspace{-0.3cm}
\end{figure*}

Interestingly, it can be found that for $i=0$, ${\bf{G}}^{-1}_{\mathrm{ZF}}={\bf{G}}^{-1}_{d,\mathrm{ZF}}=1/N\bf{I} $, and thus ${{\bf{\hat W}}_\mathrm{ZF}}=1/N{\bf{\hat H}}^{H}$, in which $1/N $ refers to the normalised factor. Therefore, we can conclude that MRC is a special case of ZF when $i=0$ in the PASCAL system, while ZF is a generalisation of MRC, which was employed in our previous work in \cite{ISAC2_Han} for SER analysis.

By using the ZF equaliser, the processed signal vector ${\bf{ x}}^{\mathrm{ZF}}_{t,l}\in \mathbb{C}{^{K \times 1}}$ can be written as 
\begin{equation}
 \setcounter{equation}{13} 
\label{x}
\!\!\!\! {\bf{ x}}^{\mathrm{ZF}}_{t,l}  ={{\bf{\hat W}}_\mathrm{ZF}} {\bf{A}}{{\boldsymbol{ \omega}}(l)}  {\bf{s}}_{t,l}\!\!+ \!\!{{\bf{\hat W}}_\mathrm{ZF}}{\bf{ n}} .
\end{equation}

The $k$th row of ${\bf{ x}}^{\mathrm{ZF}}_{t,l}$ in \eqref{x} is written as
\begin{equation}
	\label{received signal}
	{{ x}}^{\mathrm{ZF}}_{t,l,k}\!\! = \!\!\sqrt{{{P}}_{K}}{{\bf{\hat w}}_{\mathrm{ZF},k}} {{\boldsymbol{ {h } }_k}} {{s_{t,l,k}}}\!+\!\!\!\!\!\!\sum\limits_{i = 1,i \ne k}^K \!\!\!\!\!\!\sqrt{{{P}}_{i}}{{\bf{\hat w}}_{\mathrm{ZF},k}} {{\boldsymbol{ {h } }_{i}}} {{s_{t,l,{i}}}}\!+ \!{{\bf{\hat w}}_{\mathrm{ZF},k}}{\bf{ n}},
\end{equation}
where the first and second items on the right side of the equation indicate the desired and interference signals from drone $k$ and other drones, respectively. The last item indicates the noise effect.  ${{\bf{\hat w}}_{\mathrm{ZF},k}} \in \mathbb{C}{^{1 \times N}}$ indicates the vector corresponding to the $k$th row of ${{\bf{\hat W}}_{\mathrm{ZF}}}$. ${{\boldsymbol{ {h } }_k}} $ and ${{\boldsymbol{ {h } }_{i}}} $ represent the actual channel vector corresponding to drone $k$ and drone ${i}$, respectively. They can be denoted using a general expression as $\displaystyle {\boldsymbol{ {h } }_p}= \eta_p{{\bf{ a}}}({ \theta _p}) {\exp\{j2\pi { f_{D,p}}/{f_s}}\}$, in which $p \in\{1,...,K\}$. 

\subsection{Data Detection with MMSE Equalizer}
While ZF is highly efficient in eliminating interference, it suffers from noise amplification. On the contrary, MMSE achieves superior performance in minimising the mean square error by considering both interference and noise. By using the MMSE equalizer, the received signal ${{\bf{ y}}_{t,l}}$ should be multiplied by ${{\bf{\hat W}}_\mathrm{MMSE}} \in \mathbb{C}{^{K \times N}}$ as ${\bf{ x}}^{\mathrm{MMSE}}_{t,l} = {{\bf{\hat W}}_\mathrm{MMSE}}{{\bf{ y}}_{t,l}}$, in which ${{\bf{\hat W}}_\mathrm{MMSE}}$ is selected to minimize the MSE as 
\begin{align}
    \mathrm{J}({{\bf{\hat W}}_\mathrm{MMSE}})={\mathbb{E}}[||{\bf{s}}_{t,l}-{{\bf{\hat W}}_\mathrm{MMSE}}{{\bf{ y}}_{t,l}}||^2_2],
\end{align}
where the expectation is evaluated with respect to the signal and noise statistics. By using the orthogonality principle (i.e., ${\mathbb{E}}[({\bf{s}}_{t,l}-{{\bf{\hat W}}_\mathrm{MMSE}}{{\bf{ y}}_{t,l}}){{\bf{ y}}^H_{t,l}}]=0$) and considering that the channel information can be obtained by using the estimated location parameters, the optimum matrix ${{\bf{\hat W}}_\mathrm{MMSE}}$ can be obtained, which can be given by 
\begin{align}
   {{\bf{\hat W}}_{\mathrm{MMSE}}}=\mathrm{R}_{sy}\mathrm{R}^{-1}_{yy},
\end{align}
where $\mathrm{R}_{yy}={\bf{ \hat H}}\mathrm{R}_{ss} {\bf{\hat  H}}^H+\sigma^2_n\bf{I}$ and  $\mathrm{R}_{sy}=\mathrm{R}_{ss}{\bf{ \hat H}}^H$. $\mathrm{R}_{ss}$ can be written as $\mathrm{R}_{ss}={\mathbb{E}}[{\bf{s}}_{t,l}{\bf{s}}^H_{t,l}]$. Assuming that the transmitted symbols are mutually uncorrelated and have equal power, which is a commonly used assumption employed in the multiple input multiple output (MIMO) literature (e.g., \cite{MMSE}), $\mathrm{R}_{ss}=\sigma_s^2\bf{I}$ can be obtained. 

By using the Woodbury matrix identity, ${{\bf{\hat W}}_{\mathrm{MMSE}}}$ has the equivalent form as (see e.g., \cite{MMSE})
\begin{equation}
\label{W_MMSE}
    {{\bf{\hat W}}_{\mathrm{MMSE}}}=({\bf{ \hat H}}^H {\bf{ \hat H}}+\sigma^2_n/\sigma^2_s{\bf{I}})^{-1}{\bf{ \hat H}}^H.
\end{equation}

To solve the computational complexity issue of the inverse calculation for ${\bf{ \hat H}}^H {\bf{ \hat H}}+\sigma^2_n/\sigma^2_s{\bf{I}}$, the Neumann expansion is employed again. By defining ${\bf{G}}_{\mathrm{MMSE}}={\bf{ \hat H}}^H {\bf{ \hat H}}+\sigma^2_n/\sigma^2_s{\bf{I}}$, ${\bf{G}}_{\mathrm{MMSE}}$ can be approximated by using the $L$th Neumann expansion as 
\begin{align}
\label{Neumann series MMSE}
{\bf{G}}^{-1}_{\mathrm{MMSE}}&\approx\sum\limits_{{{r}} = 0}^{R}  (-{\bf{G}}^{-1}_{d,\mathrm{MMSE}}{\bf{G}}_{e,\mathrm{MMSE}})^{r}{\bf{G}}^{-1}_{d,\mathrm{MMSE}},
\end{align}
where ${\bf{G}}_{d,\mathrm{MMSE}}=(N+\sigma^2_n/\sigma^2_s)\bf{I}$ and ${\bf{G}}_{e,\mathrm{MMSE}}$ represent the matrices consisting of diagonal elements and off-diagonal elements. 

Afterwards, ${{\bf{\hat W}}_\mathrm{MMSE}}$ can also be given by 
\begin{align}
\label{Neumann series MMSE2}
    {{\bf{\hat W}}_\mathrm{MMSE}}\!\approx\!\sum\limits_{{{r}} = 0}^{R}  (-{\bf{G}}^{-1}_{d,\mathrm{MMSE}}{\bf{G}}_{\mathrm{MMSE}}+{\bf{I}})^r{\bf{G}}^{-1}_{d,\mathrm{MMSE}}{\bf{\hat H}}^{H}.
\end{align}

By using the binomial theorem, ${{\bf{\hat W}}_\mathrm{MMSE}}$ is simplified to 
\begin{align}
   {{\bf{\hat W}}_\mathrm{MMSE}}=(-1)^r\sum\limits_{{{r}} = 0}^{R}\sum\limits_{{{\tilde k}} = 0}^{r}{\left(\! {\begin{array}{*{20}{c}}
\!\!{{r}}\!\!\!\\
\!\!{{\tilde k}}\!\!\!
\end{array}} \!\right)}\frac{(-{\bf{I}})^{r-\tilde k}{\bf{G}}^{\tilde k}_\mathrm{MMSE}}{(N+\sigma^2_n/\sigma^2_s)^{\tilde k+1}}{\bf{\hat H}}^{H},
\end{align}

The $k$th row of the signal vector ${\bf{ x}}^\mathrm{MMSE}_{t,l}$, which corresponds to the MMSE-processed signal transmitted from the $k$th drone, can be expressed as 
\begin{align}
	{{ x}}^\mathrm{MMSE}_{t,l,k}\!\! &= \!\!\sqrt{{{P}}_{K}}{{\bf{\hat w}}_{\mathrm{MMSE},k}} {{\boldsymbol{ {h } }_k}} {{s_{t,l,k}}}\!+\!\!\!\!\sum\limits_{i = 1,i \ne k}^K \!\!\!\!\sqrt{{{P}}_{i}}{{\bf{\hat w}}_{\mathrm{MMSE},k}} {{\boldsymbol{ {h } }_{i}}} {{s_{t,l,{i}}}}\! \nonumber \\
    &+ \!{{\bf{\hat w}}_{\mathrm{MMSE},k}}{\bf{ n}}.
\end{align}
where ${{\bf{\hat w}}_{\mathrm{MMSE},k}}$ indicates the $k$th column of ${{\bf{\hat W}}_{\mathrm{MMSE}}}$.

\begin{figure*}[!b]
\hrulefill
 \setcounter{equation}{31} 
 \vspace{-0.2cm}
  \begin{equation}
\begin{array}{*{20}{l}}
 \label{gammax2}
\!\!\!\!\displaystyle {{ {{\Gamma }}}_\mathcal{D}}\!\!=\!\!\sum\limits_{ n_1 = 1}^N (-1)^{r_1+r_2}\sum\limits_{{{r_1}} = 0}^{R}\sum\limits_{{{r_2}} = 0}^{R}\sum\limits_{{{\tilde k_1}} = 0}^{r_1}{\left(\! {\begin{array}{*{20}{c}}
\!\!{{r_1}}\!\!\!\\
\!\!{{\tilde k_1}}\!\!\!
\end{array}} \right)}\sum\limits_{{{\tilde k_2}} = 0}^{r_2}{\left(\! {\begin{array}{*{20}{c}}
\!\!{{r_2}}\!\!\!\\
\!\!{{\tilde k_2}}\!\!\!
\end{array}} \right)} \frac{(-1)^{r_1+r_2-\tilde k_1-\tilde k_2}\sigma^2}{(C_\mathcal{D})^{\tilde k_1+\tilde k_2+2}}\sum\limits_{ i_1,...,i_{\tilde k_1+\tilde k_2}\in S_1}\sum\limits_{ q_1,...,q_{\tilde k_1+\tilde k_2}\in S_2}\prod\limits_{a_1=1}^{\tilde k_1-1}\prod\limits_{a_2=\tilde k_1+1}^{\tilde k_1+\tilde k_2-1}\ \vspace{0.3cm} \\ \!\!\!\!\ \ \ 
\times\exp\{\smash{  j\color{blue}2\pi\color{black}\big\{{ \scriptstyle \frac{ d_0}{\lambda}[{(i_{\tilde k_1+1}-i_1) \sin ({\theta _k} + \Delta {\theta _k}) + (i_{\tilde k_1}-i_{n_1})\sin ({\theta _{q_{\tilde k_1}}+\Delta \theta _{q_{\tilde k_1}}})-(i_{\tilde k_1+\tilde k_2}+i_{n_1})\sin ({\theta _{q_{\tilde k_1+\tilde k_2}}+\Delta \theta _{q_{\tilde k_1+\tilde k_2}}})-(i_{a_1+1}-i_{a_1})\sin {(\theta _{q_{a_1}}+\Delta \theta _{q_{a_1}})}}]}\big\}}\},
 \vspace{-0.8cm}
\end{array}  
 \end{equation}
 \vspace{-0.3cm}
\end{figure*}

\section{SER Analysis for PASCAL}
\label{performance analysis DTDD}

\subsection{SERs with ZF and MMSE}
\label{conditional SER MPSK}
Due to the fact that ${\bf{\hat H}}=\hat {\bf{A}}{{\boldsymbol{\hat \omega}}(l)}$ is composed of multiple random parameters (i.e., ${\hat \theta _k}$, $\hat d_k$ and ${\hat f_{D,k}} $), each of which follow the Gaussian distribution, it is not straightforward to determine the distribution of the estimated channel ${\bf{\hat H}}$. It is worth mentioning that the commonly used model for the estimated channel matrix, that is, the complex Gaussian distribution, is not suitable for our case. As a consequence, the derivation method for SER in the PASCAL system is different from that of the traditional method based on the complex Gaussian assumption. In addition, it can be noted that $\Delta { \theta _k}$ is contained in different antenna elements of the channel matrix ${\bf{\hat H}}\in\mathbb{C}{^{N \times K}}$ and the estimation errors ($\Delta { \theta _k}$, $\Delta { d_{k}}$ and $\Delta { f_{D,k}} $) are contained in both the desired and interference signals of ${{ x}}_{t,l,k}$, which result in a significant correlation issue. The product and inverse operations of ${\bf{\hat H}}$, which are caused by ZF and MMSE equalisers, further increase the complexity of the problem. 

To solve this challenging problem, we first derive the conditional SER given the estimation errors of location parameters. Thereafter, the average SER can be obtained by averaging the conditional SER over the distribution of $\Delta {\boldsymbol{  \psi}}$, in which $\Delta {\boldsymbol{  \psi}} = {[{\Delta {\boldsymbol{ \theta }}},{\Delta{\boldsymbol{ d}}},{\Delta{\boldsymbol{ f}}_{D}}]^T}$ denotes the estimation errors for location parameters. The conditional SER can be given by
\begin{equation}
\label{conditional SER}
    {P_{{e, \mathcal{D}}}|\Delta {\boldsymbol{  \psi}}}  \!\!=\!\!\sum\limits_{{m_1},...,{m_K} \in {S}}\!\! \frac{  P_{{e, \mathcal{D}}}|\{\Delta {\boldsymbol{  \psi}},\beta\} }{M^{K}} ,
\end{equation}
where $\mathcal{D}=\mathrm{ZF}$ or $\mathcal{D}=\mathrm{MMSE}$ depends on the selected equalizer. $S\in \{1,2,...,M\}$ and $M$ refers to the modulation order in MPSK. $\beta=\{{s_{t,l,1}} = {{\bf{s}}_{t,l,1}}(m_1),...,{s_{t,l,K}} = {{\bf{s}}_{t,l,K}}(m_K) \} $ indicates the phase-shift signals' combination, in which drone $1$ is transmitting ${{\bf{s}}_{t,l,1}}(m_1)$,..., drone $K$ is transmitting ${{\bf{s}}_{t,l,K}}(m_K)$, in which $m_k\in \{1,2,...,M\}$. ${{\bf{s}}_{t,l,k}}$ $\forall k\in\{1,...K\}$ denotes the $t$th signal vector in subframe $l$ from drone $k$, which is given by ${{\bf{s}}_{t,l,k}} = [1,...,{{\rm{e}}^{{{j2\pi (M - 1)}}/{M}}}]$ and thus it contains all possible phases generated by MPSK. 

\begin{figure*}[!b]
\hrulefill
 \setcounter{equation}{40} 
  \begin{align}
  \label{product-to-sum}
    &\prod_{g_1=1}^{G_1}  \sin{ x_{g_1}} \prod_{g_2=1}^{G_2}\sin{ x_{g_2}} \left(\prod_{g_3=1}^{G_3} \cos{ x_{g_3}} \prod_{g_4=1}^{G_4} \cos{ x_{g_4}}\right)^{b_1}
 \!\!\!\!= \!\! \displaystyle \!\!\!\!\!\!\sum\limits_{{e_1,...,e_{G_1\!+\!G_2\!+\!(G_3\!+\!G_4)^{b_1}}}\in S_3}\!\!\!\!\!\!\!\!\!\!\!\!\!\!\!\!\frac{(-1)^{\frac{G_1+G_2+(-1)^{b2}}{2}}f_{b3}\left(\sum\limits_{g = 1}^{G_1\!+\!G_2\!+\!(G_3\!+\!G_4)^{b_1}}   {e_g}{x_{g}}\right)\prod\limits_{g = 1}^{G_1\!+\!G_2}e_g }{2^{G_1\!+\!G_2\!+\!(G_3\!+\!G_4)^{b_1}}},
  \vspace{-0.2cm}
 \end{align}
 \vspace{-0.6cm}
\end{figure*}

When the conditional SER is considered, the only variable in the received signal is noise ${\bf{ n}}$ since the localization errors are given. As a consequence, the conditional PDF of ${{ x}}^{\mathrm{ZF}}_{t,l,k}$ and ${{ x}}^\mathrm{MMSE}_{t,l,k}$ can be given by using a general expression as 
\begin{equation}
 \setcounter{equation}{21} 
\label{conditional PDF}
\!\!\!\!\!\!\!\! \displaystyle f({{ x}}^{\mathcal{D}}_{t,l,k}|\Delta {\boldsymbol{  \psi}} ) \!=\!\frac{{\mathrm{e}}^{ -    {{{({{ x}}^{\mathcal{D}}_{t,l,k} - {\mu_\mathcal{D}} )^*}}{{{{\Gamma^{-1}_\mathcal{D} }}}}({{ x}}^{\mathcal{D}}_{t,l,k} - {\mu_\mathcal{D}} )}}}{\pi{ {{\Gamma_\mathcal{D} }}}}, \color{black}
\end{equation}
where $\mathcal{D}\in \{\mathrm{ZF}, \mathrm{MMSE} \}$ and ${\mu_\mathcal{D}} \in \{ \mu_{\mathrm{ZF}}, \mu_{\mathrm{MMSE}} \}$, in which  
\begin{align}
\mu_{\mathrm{\mathcal{D}}}\!\!=\!\!\sqrt{{{P}}_{K}}{{\bf{\hat w}}_{\mathrm{\mathcal{D}},k}} {{\boldsymbol{ {h } }_k}} {{s_{t,l,k}}}\!+\!\!\!\!\!\!\sum\limits_{i = 1,i \ne k}^K \!\!\!\!\!\!\sqrt{{{P}}_{i}}{{\bf{\hat w}}_{\mathrm{\mathcal{D}},k}} {{\boldsymbol{ {h } }_{i}}} {{s_{t,l,{i}}}}. 
\end{align}

The variance value ${{\Gamma_\mathcal{D} }}$ in \eqref{conditional PDF} can be denoted by ${\Gamma_\mathcal{D}} \in \{ \Gamma_{\mathrm{\mathcal{ZF}}}, \Gamma_{\mathrm{MMSE}} \}$, in which 
\begin{align}
    {{ {{\Gamma }}}_{\mathrm{\mathcal{D}}}} &= {\mathbb{E}}[({{ x}}^{\mathrm{\mathcal{D}}}_{t,l,k}-{\mu}_{\mathrm{\mathcal{D}}})^*({{ x}}^{\mathrm{\mathcal{D}}}_{t,l,k}-{\mu}_{\mathrm{\mathcal{D}}})] \nonumber \\
    &=\sum\limits_{ n_1 = 1}^N{{{\hat w}}^*_{\mathrm{\mathcal{D}},k,\color{blue}n_1\color{black}}}{{{\hat w}}_{\mathrm{\mathcal{D}},k,n_1}}\sigma^2,
\end{align}
where ${{{\hat w}}_{\mathrm{\mathcal{D}},k,n_1}}$ indicates the $(k,n_1)$th element of ${{\bf{\hat W}}_\mathrm{\mathcal{D}}}$.

By using the conditional PDF in \eqref{conditional PDF} and the union bound method in \cite[Sec. V-B]{ISAC2_Han}, the $  P_{{e}}|\{\Delta {\boldsymbol{  \psi}},\beta\} $ can be derived, which is written as 
\begin{align}
\setcounter{equation}{23} 
 \label{conditional SER1}
    \!\!\!P_{{e, \mathcal{D}}}|\{\Delta {\boldsymbol{  \psi}},\beta\}\!\!=\!\!\frac{1}{{\sqrt {\pi \Gamma_\mathcal{D}} }}\!\!\left(\int_{ \tilde  d_1  }^\infty \!\!\!\!{{\exp\Big\{\!\frac{{ - {{{z}  }^2}}}{\Gamma_\mathcal{D}}\!\Big \}}} d{z}\!+\!\!\!\int_{\tilde  d_2  }^\infty\!\!\!\! {{\exp\Big\{\!\frac{{ - {{{z}  }^2}}}{{\Gamma_\mathcal{D}}}}\!\Big\}} d{z}\right).
\end{align}

$  P_{{e}}|\{\Delta {\boldsymbol{  \psi}},\beta\} $ can be denoted by using the form of Q functions as
\begin{align}
     P_{{e}}|\{\Delta {\boldsymbol{  \psi}},\beta\} \!\!=\!\!Q\!\left(\! \frac {\sqrt{2 }  d_{1,\mathcal{D}}}{\sqrt{\Gamma_\mathcal{D}}} \!\right)\!+\!Q\!\left(\! \frac {\sqrt{2 }  d_{2,\mathcal{D}}}{\sqrt{\Gamma_\mathcal{D}}}\! \right)\!,
\end{align}
where  $ d_{1,\mathcal{D}}\in \{ d_{1,\mathrm{ZF}}, d_{1,\mathrm{MMSE}}  \}$ and $d_{2, \mathcal{D}} \in \{ d_{2,\mathrm{ZF}}, d_{2,\mathrm{MMSE}}  \}$. $d_{1,\mathcal{D}}$ and $d_{2,\mathcal{D}}$ can be denoted by using a general expression as $d_{\ell,\mathcal{D}}$, which can be expressed as 
    \begin{equation}
    \label{d_ell}
    d_{\ell, \mathrm{\mathcal{D}}} = \sin \left(\frac{\pi }{M} +(-1)^{\ell} \arg (\nu_\mathrm{\mathcal{D}} )\right)|\nu_\mathrm{\mathcal{D}} |,
    \end{equation}
where $\nu_\mathrm{\mathcal{D}} $ can be given by using
    \begin{align}
    \nu_{\mathrm{\mathcal{D}}}&=\sum\limits_{ p = 1}^K \!\!\sqrt{{{P}}_{p}}{{\bf{\hat w }}_{\mathrm{\mathcal{D}},k}} {{\boldsymbol{ {h } }_p}} {{{\bf{s}}_{t,l,p}}}(m_p). 
    \end{align}

By using the $R$th order Neumann series, $\nu_\mathcal{D}$ can be approximated as 
\begin{align}
 \setcounter{equation}{27} 
\label{approximate nu}
    \nu_\mathcal{D} &\approx (-1)^r \sum\limits_{{{r}} = 0}^{R}\sum\limits_{{{\tilde k}} = 0}^{r}{\left( {\begin{array}{*{20}{c}}
\!\!{{r}}\!\!\!\\
\!\!{{\tilde k}}\!\!\!
\end{array}} \right)}\frac{[(-{\bf{I}})^{r-\tilde k}{\bf{G}}^{\tilde k}_\mathcal{D}{\bf{\hat H}}^{H}]_k}{(C_\mathcal{D})^{\tilde k+1}} \nonumber 
\\
&\times  \sum\limits_{ p = 1}^K\sqrt{{{P}}_{p}}{{\boldsymbol{ {h } }_p}} {{{\bf{s}}_{t,l,p}}}
(m_p)
\end{align}
where $C_\mathrm{ZF}=N$ and  $C_\mathrm{MMSE}=N+\sigma^2_n/\sigma^2_s$. $[\cdot]_k$ represents the $k$th column of the corresponding matrix. ${\bf{G}}^{\tilde k}_\mathcal{D}$ can be denoted by using
\begin{align}
    {\bf{G}}^{\tilde k}_{\mathcal{D}}=\sum\limits_{{{v}} = 0}^{\tilde k}{\left( {\begin{array}{*{20}{c}}
\!\!{{\tilde k}}\!\!\!\\
\!\!{{v}}\!\!\!
\end{array}} \right)}(\hat {\bf{H}}^H\hat {\bf{H}})^{v}(\alpha{\bf{I}})^{\tilde k-v}, \quad
\alpha =
\begin{cases}
0, & \text{ZF},\\[4pt]
\dfrac{\sigma_n^2}{\sigma_s^2}, & \text{MMSE}.
\end{cases}
\end{align}

By substituting the value of ${\bf{G}}^{\tilde k}_\mathcal{D}$ and ${\bf{\hat H}}$ into \eqref{approximate nu} and then conducting some mathematical calculations, $\nu_\mathcal{D}$ can also be written as \eqref{nu} shown at the bottom of page \pageref{nu}, where $S_1 \in \{1,...,N\}$ and $S_2 \in \{1,...,K\}$. 

Similarly, ${{\Gamma }_\mathcal{D}}$ can be approximated as 
\begin{align}
 \setcounter{equation}{30} 
\label{gammax}
    {{ {{\Gamma }}}_\mathcal{D}} \!\!\approx\!\! \sum\limits_{ n_1 = 1}^N\left|(-1)^r \sum\limits_{{{r}} = 0}^{R}\sum\limits_{{{\tilde k}} = 0}^{r}{\left(\! {\begin{array}{*{20}{c}}
\!\!{{r}}\!\!\!\\
\!\!{{\tilde k}}\!\!\!
\end{array}} \right)}\frac{[(-{\bf{I}})^{r-\tilde k}{\bf{G}}^{\tilde k}_\mathcal{D}{\bf{\hat H}}^{H}]_{k,n_1}}{(C_\mathcal{D})^{\tilde k+1}}\!\right|^2\!\sigma^2,
\end{align}
where $[\cdot] _{k,n_1}$ represents the $(k,n_1)$th column of the corresponding matrix. By substituting the value of ${\bf{G}}^{\tilde k}_\mathcal{D}$ and ${\bf{\hat H}}$ into \eqref{gammax} and then conducting some mathematical calculations, ${{ {{\Gamma }}}_\mathcal{D}}$ can also be written as \eqref{gammax2} shown at the bottom of page \pageref{gammax2}. 

The average SER can be obtained by averaging the conditional SER over the distribution of $\Delta {\boldsymbol{  \psi}}$ as 
\begin{align}
\setcounter{equation}{32} 
\label{average SER}
  {P_{{e,\mathcal{D}}}}&= \sum\limits_{{m_1},...,{m_K} \in {S}} \frac{  {\mathbb{E}}[P_{{e,\mathcal{D}}}|\{\Delta {\boldsymbol{  \psi}},\beta\}] }{M^{K}} \nonumber
  \\&= \sum\limits_{{m_1},...,{m_K} \in {S}} \frac{1}{M^{K}}\left(  {\mathbb{E}}[Q_{1,\mathcal{D}} ]+{\mathbb{E}}[Q_{2,\mathcal{D}} ]\right),
\end{align}
where $Q_{1,\mathcal{D}}=Q\!\left(\! \frac {\sqrt{2 } d_{1, \mathcal{D}}}{\sqrt{  \Gamma_\mathcal{D}}} \!\right)\!$ and $Q_{2,\mathcal{D}}=\!Q\!\left(\! \frac {\sqrt{2 } d_{2, \mathcal{D}}}{\sqrt{\Gamma_\mathcal{D}}}\! \right)$. \vspace{0.2cm}

The analytical derivation of average SERs is inherently challenging owing to the following factors:
\begin{enumerate}
    \item The localization errors for all drones exert an influence on the average SER as can be observed in $\nu_\mathcal{D}$ in \eqref{nu} and ${{ {{\Gamma }}}_\mathcal{D}}$ in \eqref{gammax2}. 
    \item The numerator $d_{\ell, \mathcal{D}}$ and the denominator $\Gamma_\mathcal{D}$ contain the same random variables (i.e., $\Delta {\theta}$, $\Delta {d}$ and $\Delta {f_D}$), which leads to a strong correlation between them. 
    \item The Q function comprises a complex function involving random variables, for which a direct expectation evaluation does not yield a closed-form solution.
    \item Even though the $R$th Neumann series provides a feasible solution for the inverse calculation, the computational complexity is still high.
\end{enumerate}

\subsection{The analytical derivation for Average SERs}
\label{Approximation Methods}

To begin, we convert $\arg(\cdot)$ in \eqref{d_ell} into the form of $\arctan(\cdot)$ according to the values of $\nu_x$ and $\nu_y$, which represent the real and image parts of $\nu_\mathcal{D}$, respectively. Thereafter, by using $\cos (\arctan (x)) = \frac{1}{{\sqrt {1 + {x^2}} }}$ and $\sin (\arctan (x)) = \frac{x}{{\sqrt {1 + {x^2}} }} $ and conducting some mathmetical operations, ${\mathbb{E}}[Q_{\ell},_\mathcal{D} ]$ for ${\ell}\in \{1,2\} $ in \eqref{average SER} can also be written as 
    \begin{equation}
    \label{EQ_ell}
{\mathbb{E}}[Q_{{\ell}, \mathcal{D}}]  \!\!= \!\!
\begin{cases}
  {\mathbb{E}}\left[Q\left( \frac {\sqrt{2 } \tilde \nu}{\sqrt{{{{{\Gamma }}}_\mathcal{D}}}} \right)\right],  
  & \text{if } \nu_{x}  \neq  0 \vspace{0.1cm} \\
  {\mathbb{E}}\left[Q \left( {\frac{ \sqrt{2} C_1\nu_{y}}{\sqrt{{{{\Gamma }}}_\mathcal{D}}}} \right)\right], & \text{if } \nu_{x} = 0\! \text{ and }\! \nu_{y}\neq 0   \vspace{0.1cm} \\
  \text{undefined}, & \text{if } \nu_{x} = 0\! \text{ and }\! \nu_{y} = 0 \\
\end{cases} 
\end{equation}
where $\tilde \nu=\sin \frac{\pi }{M} \nu _{x} + (-1)^{\ell}\cos \frac{\pi }{M}\nu _{y}$ and $C_1=(-1)^{{\ell}}{\cos\! \frac{\pi }{M}}$.

\begin{figure*}[!b]
\hrulefill
 \setcounter{equation}{42} 
 \vspace{-0.2cm}
\begin{align}
\label{E_7}
   {\mathbb{E}}_{\ell+6}\!\!=\!\! &\sum\limits_{{ G_2}=0}^{G_1\!+\!G_2}\! {\left(\! {\begin{array}{*{20}{c}}
\!\!\!{{G_1\!+\!G_2}}\!\!\!\!\\
\!\!\!{{G_2}}\!\!\!\!
\end{array}} \!\right)} \! \left(\sum\limits_{{ G_4}=0}^{G_3\!+\!G_4} \! {\left(\! {\begin{array}{*{20}{c}}
\!\!\!{{G_3\!+\!G_4}}\!\!\!\\
\!\!\!{{G_4}}\!\!\!
\end{array}} \!\right)} \!\! \right )^{b_1} \!\!\frac{(-1)^{G_2\!+\!(G_4)^{b_1}} 2^{{G_1\!+\!(G_3)^{b_1}}}   }{N^{{\tilde G}_2}}  \!\!\!\!\!\!   \sum\limits_{ i_1,...,i_{{\tilde G}_1} \in  S_1}  \sum\limits_{j_1,...,j_{G_2\!+\!(G_4)^{b_1}} \in  S_1}   \sum\limits_{ q_1,...,q_{G_2\!+\!(G_4)^{b_1}}  \in  S_2}\sum\limits_{  p_1,...,p_{{\tilde G}_1}  \in  S_2} \nonumber \\
&{\prod\limits_{g = 1}^{{\tilde G}_1}\eta_g \sqrt{P_{p_g}}} \sum\limits_{{e_1,...,e_{{\tilde G}_1}}\in S_3} \frac{(-1)^{\frac{G_1\!+\!G_2\!+(-1)^{b_2}}{2}}\prod\limits_{g = 1}^{G_1\!+\!G_2\!}e_g}{2^{{\tilde G}_1}}\underbrace{{\mathbb{E}}[f_{b_3}(C_1+C_2\Delta f_{D,k}+C_3\Delta \theta_{k}+C_4\Delta \theta_{q_g})]}_{{\mathbb{E}}_9}
\end{align}
 \vspace{-0.3cm}
\begin{align}
 \setcounter{equation}{44} 
\label{C_1}
   {{C}}_1\!\!&=\!\! \sum\limits_{ g = 1}^{{\tilde G}_1} 2 \pi e_g \! \left(\! \textstyle{ \frac{m_{p_g}-1}{M}\!+\!\frac{f_{D,p_g}-f_{D,k}}{f_s}}\!\right)\!+\!\!\!\!\!\!\!\sum\limits_{ g = 1}^{ G_1\!+\!(G_3)^{b_1}}\!\!\!\!\!\!2 \pi e_g \!\left(\! \textstyle{ d_0\frac{(i_g-1)(\sin {\theta _k}  - \sin {\theta _p})}{\lambda}}\!\right)\!+\!\!\!\!\!\!\sum\limits_{ g = 1}^{G_2\!+\!(G_4)^{b_1}}\!\!\!\!\!2 \pi e_g \!\left(\! \textstyle{d_0\frac{(i_g-1) \sin {\theta _k}  - (j_g-1)\sin {\theta _{p_g}}+(j_g-i_g)\sin {\theta _{q_g}}}{\lambda} }\!\right).
\end{align}
\vspace{-0.3cm}
\end{figure*}

Since there is no closed-form expression for ${\mathbb{E}}_1$ (i.e., ${\mathbb{E}} [Q_{{1}, \mathcal{D}}]$ ) and ${\mathbb{E}}_2$ (i.e., ${\mathbb{E}} [Q_{{2}, \mathcal{D}}]$ ), the Taylor approximation proposed is invoked, which is given by 
\begin{equation}
\setcounter{equation}{35} 
\label{Taylor approximation}
    Q(x)\simeq\sum\limits_{{r} = 0}^{{R}} {\frac{Q^{(r)}(x_0){{{(x- {x_{0}})}^{{r}}}}}{{{r}!}}},
    \vspace{-0.1cm}
\end{equation}
where $Q^{(r)}(x_0)$ represents the $r$th derivative of the Q function at $x_0$ and $Q^{(0)}(x_0)=Q(x_0)$. For $r>0$, $Q^{(r)}(x_0)$ can be written as
\begin{equation}
    Q^{(r)}(x_0)=-\frac{1}{(\sqrt{2})^{r}\sqrt{\pi}}{\rm{e}}^{-\frac{x^2_0}{2}}(-1)^{r+1}H_{r-1}\big(\frac{x_0}{\sqrt{2}}\big),
\end{equation}
where $\displaystyle H_{r-1}\big(\cdot\big)$ denotes the $(r-1)$th Hermite polynomial.

By using \eqref{Taylor approximation}, the Q functions in ${\mathbb{E}}_1$ and ${\mathbb{E}}_2$ can be converted into polynomials as
 \begin{equation}
  \setcounter{equation}{36}
  \label{E_ell}
    {\mathbb{E}}_{\ell}\simeq \sum\limits_{{r} = 0}^{{R}}\! \frac{(\sqrt{2})^r Q^{(r)}(x_{{\ell},0})(C^r_1)^{\ell-1}{\mathbb{E}}_{\ell+2}}{r!},
\end{equation}  
where $ x_{\ell,0}= \frac{\sqrt{2 } \tilde \nu_{ 0}}{\sqrt{{{{{\Gamma }}}_{\mathcal{D},0}}}}$ for ${\ell}=1$ and $ x_{\ell,0}={\frac{ \sqrt{2} C_1\nu_{y,0}}{\sqrt{{{{\Gamma }}}_{\mathcal{D},0}}}}$ if ${\ell}=2$. $\vspace{0.1cm} \tilde \nu_{0}$, $\nu_{y,0}$ and ${{{\Gamma }}}_{\mathcal{D},0} $ represent the prefect case of ${\tilde \nu}$, $\nu_{y}$ and ${{{\Gamma }}}_{\mathcal{D}} $ when $\Delta {\boldsymbol{  \psi}}=0 $. Furthermore, $\vspace{0.1cm}{\mathbb{E}}_3={\mathbb{E}}[(\frac { \tilde \nu}{\sqrt{{{{{\Gamma }}}_{\mathcal{D}}}}}- \frac { \tilde \nu_{0}}{\sqrt{{{{{\Gamma }}}_{\mathcal{D},0}}}})^r]$ and ${\mathbb{E}}_4={\mathbb{E}}[({\frac{ \nu_{y}}{\sqrt{{{{\Gamma }}}_\mathcal{D}}}}- {\frac{ \nu_{y,0}}{\sqrt{{{{\Gamma }}}_{\mathcal{D},0}}}})^r] $, respectively. 

By using the binomial theorem to expand the polynomials in ${\mathbb{E}}_3$ and ${\mathbb{E}}_4$, ${\mathbb{E}}_3$ and ${\mathbb{E}}_4$ become
\vspace{-0.2cm}
\begin{equation}
\label{E 3 E 4}
 {{\mathbb{E}}_{\ell+2}} = \sum\limits_{{k_1}=0}^{r} {\left(\! {\begin{array}{*{20}{c}}
{{r}}\\
{{k_1}}
\end{array}} \!\right)}  {\mathbb{E}}_{\ell+4} C_{\ell+2}^{{{r-k_1}}},  
\vspace{-0.2cm}
\end{equation}
where ${\mathbb{E}}_{\ell+4}$ for $\ell\in\{1,2\}$ can be given by ${\mathbb{E}}_5= \displaystyle {\mathbb{E}}\Big[\! {\left(\frac {{ \tilde \nu}}{{\sqrt{{{{{\Gamma }}}_{\mathcal{D}}}}}}\right)^{{k_1}}}\!\Big]   $ and ${\mathbb{E}}_6=\displaystyle {\mathbb{E}}\Big[\! {\left(\frac {{ \nu_y}}{{\sqrt{{{{{\Gamma }}}_{\mathcal{D}}}}}}\right)^{{k_1}}}\!\Big]$, $C_3=-\frac { \tilde \nu_{ 0}}{\sqrt{{{{{\Gamma }}}_{\mathcal{D},0}}}}$ and $C_4=-{\frac{ \nu_{y,0}}{\sqrt{{{{\Gamma }}}_{\mathcal{D},0}}}} $. 

By using the first-order Taylor approximation for fractions, ${\mathbb{E}}_5$ and ${\mathbb{E}}_6$ can be approximated as ${\mathbb{E}}_5\approx \displaystyle \! {\frac {{\mathbb{E}}[({ \tilde \nu})^{{k_1}}]}{{\mathbb{E}}[\big({\sqrt{{{{{\Gamma }}}_{\mathcal{D}}}}}\big)^{{k_1}}]}}\!   $ and ${\mathbb{E}}_6\approx \displaystyle \! {\frac {{\mathbb{E}}[({  \nu_{y}})^{{k_1}}]}{{\mathbb{E}}[\big({\sqrt{{{{{\Gamma }}}_{\mathcal{D}}}}}\big)^{{k_1}}]}}\!   $, in which by using the binomial theorem,
\begin{align}
 \!\!\!\!  {{\mathbb{E}}[({ \tilde \nu})^{{k_1}}]}&\!\!=\!\!{{\mathbb{E}}[( \sin \frac{\pi }{M} \nu _{x} + (-1)^{\ell}\cos \frac{\pi }{M}\nu _{y})^{{k_1}}]}\nonumber \\&\!\!=\!\!\sum\limits_{{k_2}=0}^{k_1} {\left(\! {\begin{array}{*{20}{c}}
\!\!{{k_1}}\!\!\\
\!\!{{k_2}}\!\!
\end{array}} \!\right)}  \left(\!\sin \frac{\pi }{M}\!\right)^{k_2} \left(\!(-1)^{\ell}\cos\!\!\frac{\pi }{M}\!\right)^{k_1\!-\!k_2}{\mathbb{E}}_7, 
\end{align}
where ${\mathbb{E}}_7={\mathbb{E}}[\nu^{k_2}_y\nu^{k_1-k_2}_x]$. By calculating the cosine and sine values of $\nu$ in \eqref{approximate nu}, $\nu_x$ and $\nu_y$ can be obtained, respectively. Afterwards, by using the binomial theorem and performing some basic mathematical operations, $\nu^{k_2}_y$, $\nu^{k_1-k_2}_x$  and $\nu^{k_1}_y$ can be simplified to a general expression as 
\begin{align}
  \nu^{\tilde n}_b&=  \sum\limits_{{\tilde k}=0}^{\tilde n} {\left(\! {\begin{array}{*{20}{c}}
\!\!{{\tilde n}}\!\!\\
\!\!{{\tilde k}}\!\!
\end{array}} \!\right)} \Big(\frac{2}{N}\color{black} \sum\limits_{ p = 1}^K \sum\limits_{ i = 1}^N \eta_p\sqrt{{{P}}_{p}} f_b(2\pi(\Theta_1+\Phi_1))\Big)^{\tilde n-\tilde k}
\nonumber \\
&\times\Big(-\color{blue}\frac{1}{N^2}\color{black} \sum\limits_{ i = 1}^N\sum\limits_{ j = 1}^N \sum\limits_{ q = 1}^K\sum\limits_{ p = 1}^K \eta_{p}\sqrt{P_p} f_b(2\pi(\Theta_2+\Phi_1)) \Big)^{\tilde k},
\end{align}
where $b\in \{x,y\}$, $\tilde n \in \{k_2, k_1-k_2, k_1 \} $ and $\tilde k \in \{k_3, k_4, k_5 \}  $. $f_b (\cdot) = \cos (\cdot) $ for $b=x$, while $f_b (\cdot) = \sin (\cdot) $ for $b=y$. 

Then, we can expand brackets in $\nu^{k_2}_y$, $\nu^{k_1-k_2}_x$ and $\nu^{k_1}_y$. For instance, we use
\vspace{-0.2cm}
\begin{align}
\setcounter{equation}{39} 
  \left(\sum\limits_{ i = 1}^N \cos\theta_i\right)^{\tilde k}=\sum\limits_{ i_1,...,i_{\tilde k}\in S_1} \prod\limits_{g = 1}^{\tilde k}\cos\theta_{i_g},  
  \vspace{-0.2cm}
\end{align}
where the subscript $g$ is used to distinguish different $ \cos \theta_i $ in the product operation. Thereafter, we apply the product-to-sum identities of the trigonometric formulas to $\nu^{k_2}_y$, $\nu^{k_1-k_2}_x$ and $\nu^{k_1}_y$ to convert the products of sine and cosine functions into sum or difference forms in order to simplify the calculation of ${\mathbb{E}}_7={\mathbb{E}}[\nu^{k_2}_y\nu^{k_1-k_2}_x]$ and ${\mathbb{E}}_8={{\mathbb{E}}[({  \nu_{y}})^{{k_1}}]} $. 

The employed product-to-sum identity is shown \eqref{product-to-sum} at the bottom of page \pageref{product-to-sum}, in which $S_3 \in \{1,-1\}$. For simplifying $\nu^{k_2}_y\nu^{k_1-k_2}_x$, $b_1$ in \eqref{product-to-sum} is set to $b_1=1$, while $b_1$ in \eqref{product-to-sum} is set to $b_1=0$ to simplify ${  \nu^{{k_1}}_{y}} $. In \eqref{product-to-sum}, $b_2=0$ if \textit{$G_1$} and \textit{$G_2$} have the same parity, while $b_2=1$ if  \textit{$G_1$} and \textit{$G_2$} have the different parity. The function $f_{b_3} (\cdot)$ can be represented as 
\begin{align}
 \setcounter{equation}{41} 
 \label{fb3}
    f_{b_3} (\cdot) \!\!= \!\!
    \begin{cases}
        \cos(\cdot),   \ \ \    \text{if \textit{$G_1$} and \textit{$G_2$} have the same parity } \!\!\!\!
        \\
        \sin(\cdot),  \ \ \  \text{if  \textit{$G_1$} and \textit{$G_2$} have the distinct parity } \!\!  \!\!
    \end{cases} \!\!\!\!
\end{align}

After that, by employing the product-to-sum identity shown in \eqref{product-to-sum}, and conducting some simple algebraic operations, ${\mathbb{E}}_7$ and ${\mathbb{E}}_8$ can be simplified as \eqref{E_7} at the bottom of page \pageref{E_7}, in which $\ell\in\{1,2\}$. $\tilde G_1$ and $\tilde G_2$ can be denoted by using a general expression, which is written as
    \begin{equation}
    \setcounter{equation}{44} 
      \tilde G_\ell= G_1+(2)^{\ell-1}G_2+(G_3+(2)^{\ell-1}G_4)^{b_1} ,
    \end{equation}
where $G_1=k_3-k_2$, $G_2=k_3$, $G_3=k_1-k_2-k_4$, $G_4=k_4$ and $b_1=1$ for ${\mathbb{E}}_7$, while $G_1=k_1-k_5$, $G_2=k_5$ and $b_1=0$ for ${\mathbb{E}}_8$.

In ${\mathbb{E}}_9$ in \eqref{E_7}, $C_1$ is shown in \eqref{C_1} on page \pageref{C_1}, while $C_2$ can be given by $C_2=\sum\limits_{ g = 1}^{{\tilde G}_1} \frac{-2 \pi e_g}{f_s}$. $C_3$ and $C_4$ can be written as 
 \setcounter{equation}{45}
\begin{subequations} 
\vspace{-0.3cm}
\begin{equation}
     C_3= \sum\limits_{ g = 1}^{{\tilde G}_1} \frac{2 \pi e_g d_0(i_g-1)\cos\theta_k}{\lambda},   
\end{equation}
\vspace{-0.2cm}
\begin{equation}
    C_4=\sum\limits_{ g = 1}^{G_2+(G_4)^{b_1}} \frac{2 \pi e_g d_0(j_g-i_g)\cos\theta_{q_g}}{\lambda},
\end{equation}
\vspace{-0.3cm}
\end{subequations}

The closed-form expression of ${\mathbb{E}}_9$ is shown in Appendix \ref{Appendix_A}, which is then substituted into \eqref{E_7} to obtain ${\mathbb{E}}_{\ell+6}$. 

\begin{figure*}[!b]
\hrulefill
 \vspace{-0.2cm}
\begin{align}
  \setcounter{equation}{47} 
 \!\!\!\displaystyle \!\!&=\!\! -\!\frac{8\sigma^4}{N^7}\!\!\!\! \!\!\sum\limits_{n_1,n_2\in S_1}\!\sum\limits_{i_1,...,i_3 \in S_1}\!\sum\limits_{q_1,...,q_3 \in S_2}\!\!\!\! \eta^2_k \prod\limits_{q \in S_5}\!\!  \eta^2_{q} \underbrace{{\mathbb{E}}[\cos(\Theta_4(i_3,n_2))\exp\{j \Theta_3(i_1,i_2,n_1)\}]}_{{\mathbb{E}}_{15}} \!-\!\frac{16\sigma^4}{N^4}\!\! \! \sum\limits_{n_1=1}^{N}\!\sum\limits_{i_1=1}^{N}\!\sum\limits_{q_1=1}^{K} \eta^3_k\eta^2_{q_1} \underbrace{{\mathbb{E}}[\cos(\Theta_4(i_1,n_1))]}_{{\mathbb{E}}_{13}}  \nonumber \\&+\frac{4\sigma^2}{N^2} \eta^2_k{\mathbb{E}}[{{ {{\Gamma }}}_x}]\!+\!\color{blue}\frac{4\sigma^4}{N^5}\color{black}\sum\limits_{n_1 =1}^{N}\! \sum\limits_{i_1,i_2 \in S_1}\!\sum\limits_{q_1,q_2 \in S_2}\!\!\eta^3_k \!\prod\limits_{q \in S_4}\! \eta^2_{q}\underbrace{{\mathbb{E}}[\exp\{ {j \Theta_3(i_1,i_2,n_1)}\}]}_{{\mathbb{E}}_{11}}\!+\!\frac{\sigma^4}{N^8}\!\!\!\! \sum\limits_{n_1,n_2\in S_1}\!\sum\limits_{i_1,...,i_4 \in S_1}\!\sum\limits_{q_1,...,q_4 \in S_2}\!\!\!\!\eta^2_k \prod\limits_{q \in S_6} \!\!\eta^2_{q}  \nonumber\\  
&\times \underbrace{{\mathbb{E}}[\exp\{j \Theta_3(i_1,i_2,n_1)+\Theta_3(i_3,i_4,n_2)\}]}_{{\mathbb{E}}_{12}}+\!\frac{8\sigma^4}{N^6}\!\!\!\! \sum\limits_{n_1,n_2\in S_1}\!\sum\limits_{i_1,i_2 \in S_1}\!\sum\limits_{q_1,q_2 \in S_2} \eta^2_k \prod\limits_{q \in S_4}\eta^2_{q} \underbrace{{\mathbb{E}}[\sum\limits_{j \in \ell}\cos((\Theta_4(i_1,n_1)+(-1)^{j}\Theta_4(i_2,n_2)))]}_{{\mathbb{E}}_{14}} \vspace{0.1cm},
\label{EEE2}
\end{align}
 \vspace{-0.3cm}
\end{figure*}

In ${\mathbb{E}}_5$ and ${\mathbb{E}}_6$, since there is no closed-form expression for ${{\mathbb{E}}[\big({\sqrt{{{{{\Gamma }}}_{\mathcal{D}}}}}\big)^{{k_1}}]}$, the second-order Taylor approximation for ${{\mathbb{E}}[\big({\sqrt{{{{{\Gamma }}}_{\mathcal{D}}}}}\big)^{{k_1}}]}$ is invoked as
\begin{equation}
\setcounter{equation}{47}
\label{E_sqrt_gammax_k1}
    {{\mathbb{E}}[\big({\sqrt{{{{{\Gamma }}}_{\mathcal{D}}}}}\big)^{{k_1}}]} \approx {{\mathbb{E}}[\big({{{{{{\Gamma }}}_{\mathcal{D}}}}}\big)]}^{{\frac{k_1}{2}}}+\frac{k_1(k_1-2){{\mathbb{E}}[\big({{{{{{\Gamma }}}_{\mathcal{D}}}}}\big)]}^{{\frac{k_1}{2}-2}}{\mathbb{E}}_{10}}{2^22!},
\end{equation}
where ${\mathbb{E}}_{10}={{\mathbb{E}}[\big({{{\Gamma }}}_{\mathcal{D}}-{{\mathbb{E}}[{{{\Gamma }}}_{\mathcal{D}}]}\big)^2]}={\mathbb{E}}[{{{\Gamma }}}^2_{\mathcal{D}}]-{{\mathbb{E}}[{{{\Gamma }}}_{\mathcal{D}}]}^2$. ${{\mathbb{E}}[{{{\Gamma }}}^2_{\mathcal{D}}]}$ can be simplified by using some simple algebraic operations to \eqref{EEE2} as shown in the bottom of page \pageref{EEE2}, in which $S_4\in\{q_1,q_2\}$, $S_5\in\{q_1,q_2,q_3\}$ and $S_6\in\{q_1,q_2,q_3,q_4\}$. Since ${{\mathbb{E}}[{{{\Gamma }}}_{\mathcal{D}}]}$ and ${{\mathbb{E}}[{{{\Gamma }}}^2_{\mathcal{D}}]}$ contain the same components ( i.e., ${\mathbb{E}}_{11}={\mathbb{E}}[\exp\{j2\pi \Theta_3(i_1,i_2,n_1)\}]$ and ${\mathbb{E}}_{14}={\mathbb{E}}[\cos\{ \Theta_4(i_1,n_1)\}]$ ), we will now proceed to calculate ${{\mathbb{E}}[{{{\Gamma }}}_{\mathcal{D}}]}$ and ${{\mathbb{E}}[{{{\Gamma }}}^2_{\mathcal{D}}]}$ together. To begin, ${\mathbb{E}}_{11}$ and ${\mathbb{E}}_{12}$ can be expressed by using a general expression as
\begin{align}
\setcounter{equation}{48} 
\!\!\!\!\!\!\mathbb{E}_{\ell+10} 
\!\!=\!\! \mathbb{E}\bigg[& \exp \Bigg\{ j C_5 \Bigg(
    \sum_{c=1}^{2+2^{\ell-1}} (-1)^c i_c \sin(\theta_k + \Delta \theta_k) \notag\\
    &- \sum_{e=1}^{1+1^{\ell-1}}\!\!\!\! \sum_{c=2e-1}^{2e} (n_e - i_c) 
       \sin(\theta_{q_c} + \Delta \theta_{q_c})
\Bigg)\Bigg\}\bigg].\!\!
\end{align}
where $C_5={2\pi d_0}/{\lambda}$. By employing Euler’s formula and the small-angle approximation (i.e., $\sin ({ \theta _k}+\Delta{ \theta _{k}})
    \simeq \sin { \theta _k}+\Delta{ \theta _{k}}\cos { \theta _k}$), and performing some simple mathematical manipulations, $\mathbb{E}_{\ell+10}$ can be simplified to $\mathbb{E}_{\ell+10}=\sum\limits_{ r = 1}^2 j^{r-1}\mathbb{E}_{16}$, in which
 \begin{align}
 \label{E_16}
 \mathbb{E}_{16} 
\!\!=\!\! \mathbb{E}\left[f_r(C_6\!+\!C_7\cos \theta_k\Delta \theta_k
\!+\!C_8\cos \theta_{q_c}\Delta \theta_{q_c})\right], 
\end{align}
where $f_1=\cos(\cdot)$ and $f_2=\sin(\cdot)$. $C_6=C_7\sin \theta_k+C_8\sin \theta_{q_c}$, in which 
\begin{subequations}
\begin{align}
C_7&=\sum\limits_{c=1}^{2+2^{\ell-1}}C_5 (-1)^c i_c, \\
    C_8&=\sum\limits_{e=1}^{1+1^{\ell-1}}\!\!\!\! \sum\limits_{c=2e-1}^{2e} C_5(-1)^c(n_e - i_c).
\end{align}    
\end{subequations}

\begin{table}[t]
\small
\caption{Simulation Setup}
\vspace{-0.2cm}
\centering{}
\newsavebox{\tablebox}
\begin{lrbox}{\tablebox}
\begin{tabular}{|c|c|c|c|c|c|}
\hline
\textbf{Param.}                    &   \textbf{Value} &\textbf{Param.}                    &   \textbf{Value} &\textbf{Param.}                    &   \textbf{Value}
\\
\hline
$K$      &4    &    $T$    & 100  &  ${N_\mathrm{MC}}$      &$5 \times 10^{3}$\\
\hline
${\theta}_1$     & $10^ \circ$  & ${d}_1$     & $80 \; \mathrm{m}$  & ${f}_{D,1}$     & $ 1.0\mathrm{kHz}$ 
\\
\hline
${\theta}_2$     & $25^ \circ$  & ${d}_2$     & $80 \; \mathrm{m}$  & ${f}_{D,2}$     & $ 2.5\mathrm{kHz}$ 
\\
\hline
${\theta}_3$     & $40^ \circ$  & ${d}_3$     & $80 \; \mathrm{m}$  & ${f}_{D,3}$     & $ 4.0\mathrm{kHz}$ 
\\
\hline
${\theta}_4$     & $55^ \circ$  & ${d}_4$     & $80 \; \mathrm{m}$  & ${f}_{D,4}$     & $ 5.5\mathrm{kHz}$ 
\\
\hline
$v_1$     & $2 \; \mathrm{m/s}$  & $v_2$     & $5 \; \mathrm{m/s}$  &  $v_3$     & $8 \; \mathrm{m/s}$
\\
\hline
$v_4$     & $11 \; \mathrm{m/s}$   &   $f_s$       & 100 $\mathrm{kHz}$ &$\lambda$     & 2  $\mathrm{mm}$ 
\\
\hline
\end{tabular}
\end{lrbox}
\scalebox{0.9}{\usebox{\tablebox}}
\label{sim par}
\vspace{-0.3cm}
\end{table}

  \begin{figure*}[t]
\centering 
{\includegraphics  [height=1.96in, width=7in]{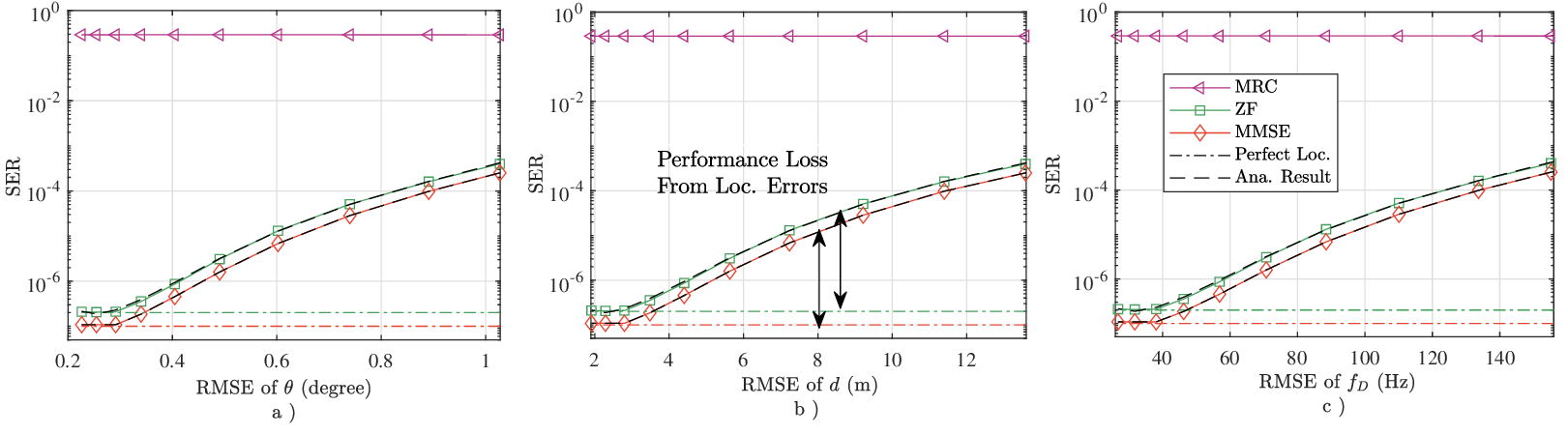}}
\vspace{-0.3cm}
\caption{ The effect of estimation errors of a) DOA $\theta$, b) Doppler frequency $f_D$, c) Range $d$ on SER} 
\label{RMSE_SER}
\end{figure*}

  \begin{figure}[ptb]
\centering
{\includegraphics  [height=2.21in, width=2.91in]{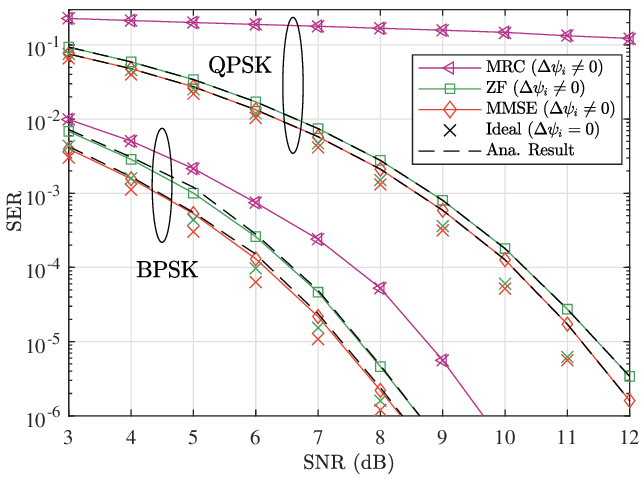}}
\vspace{-0.4cm}
\caption{SER of drone $k$ with and without the localization errors of other drones (i.e., $\Delta {\boldsymbol {\psi}}_i $ )}
\color{black}
\label{other_drones_effect}
\end{figure}

 \begin{figure}[ptb]
\centering
{\includegraphics  [height=2.21in, width=2.91in]{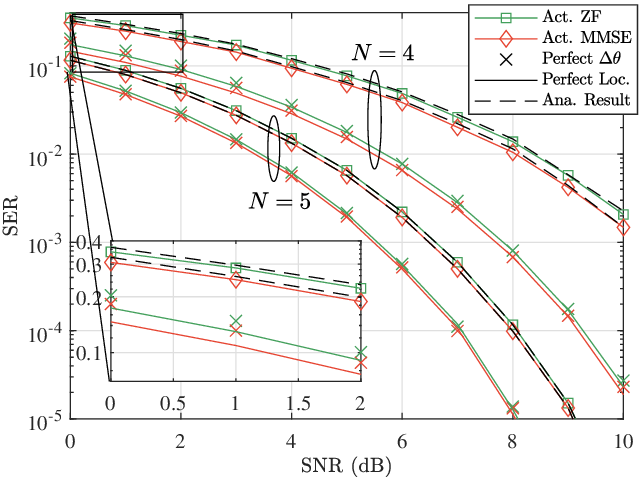}}
\vspace{-0.4cm}
\caption{SER comparison for scenarios with actual results, perfect DOA estimation (i.e., $\Delta \theta =0$), and perfect localization  errors}
\color{black}
\label{dominant_parameter}
\vspace{-0.4cm}
\end{figure}

By using the small-angle approximation and then conducting some simple mathematical manipulations again, ${{\mathbb{E}}_{13}}$ in ${{\mathbb{E}}[{{{\Gamma }}}_{\mathcal{D}}]}$ and ${{\mathbb{E}}[{{{\Gamma }}}^2_{\mathcal{D}}]}$ can be simplified to
 \begin{align}
 \label{E_13}
\mathbb{E}_{13}=\mathbb{E}\left[\cos(C_9\!+\!C_{10}\cos \theta_k\Delta \theta_k
\!+\!C_{11}\cos \theta_{q_1}\Delta \theta_{q_1})\right],
 \end{align}
where $C_{10}=C_5 (i_1-n_1) $, $C_{11}=C_5 (n_1-i_1) $ and $C_{9}=C_{10}\sin \theta_k+C_{11}\sin \theta_{q_1} $. Similarly, ${{\mathbb{E}}_{14}}$ can be simplified to ${{\mathbb{E}}_{14}}=\frac{1}{2}\sum\limits_{ r = 1}^2\!\mathbb{E}_{17}$, in which
 \begin{equation}
 \label{E_17}
\mathbb{E}_{17}\!\!=\!\!\!\mathbb{E}\!\bigg[\!\cos\!\bigg(\!C_{12,r}\!+\!C_{13,r}\cos \theta_k\Delta \theta_k
\!+\!C_{14,r}\cos \theta_{q_c}\Delta \theta_{q_c}\!\bigg)\!\bigg],      
\vspace{-0.2cm}
 \end{equation}
where $C_{12,r}=C_{13,r}\sin \theta_k+C_{14,r}\sin \theta_{q_c} $, in which
 \begin{subequations}
     \begin{align}
     C_{13,r}&=\sum\limits_{ c = 1}^2 (-1)^{r-1+c}C_5 (n_c-i_c), \\
     C_{14,r}&=(-1)^{r+c} C_{13,r}.
          \end{align}
 \end{subequations}

    \begin{figure*}[t]
\centering 
{\includegraphics  [height=1.96in, width=7in]{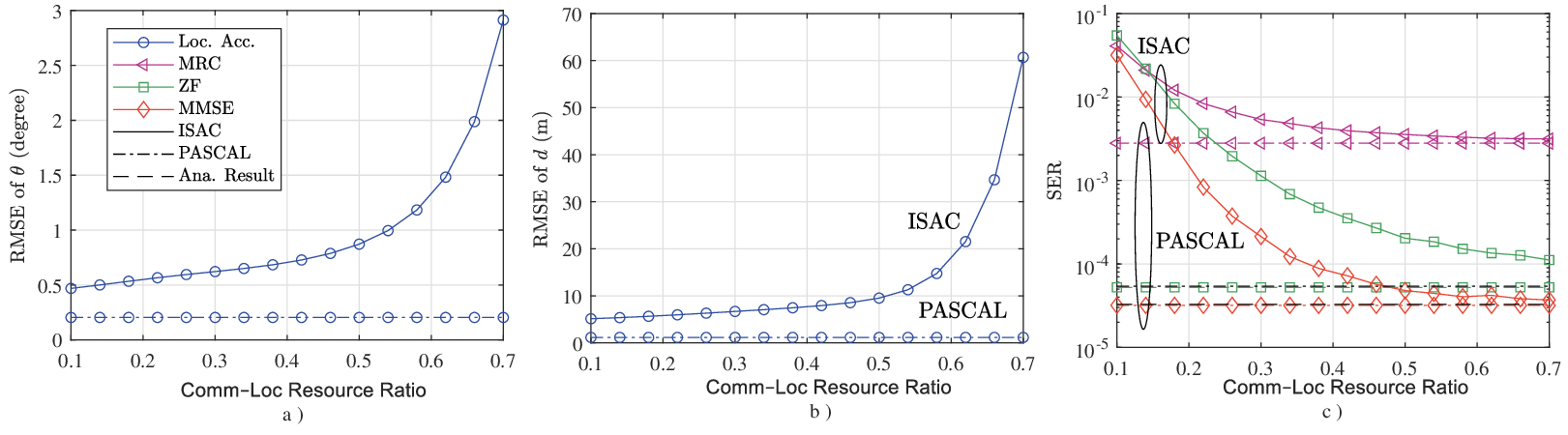}}
\vspace{-0.3cm}
\caption{ The Comparison between ISAC and PASCAL} 
\label{ISAC_PASCAL}
\end{figure*}

\begin{figure*}[t]
\centering 
{\includegraphics  [height=1.96in, width=7in]{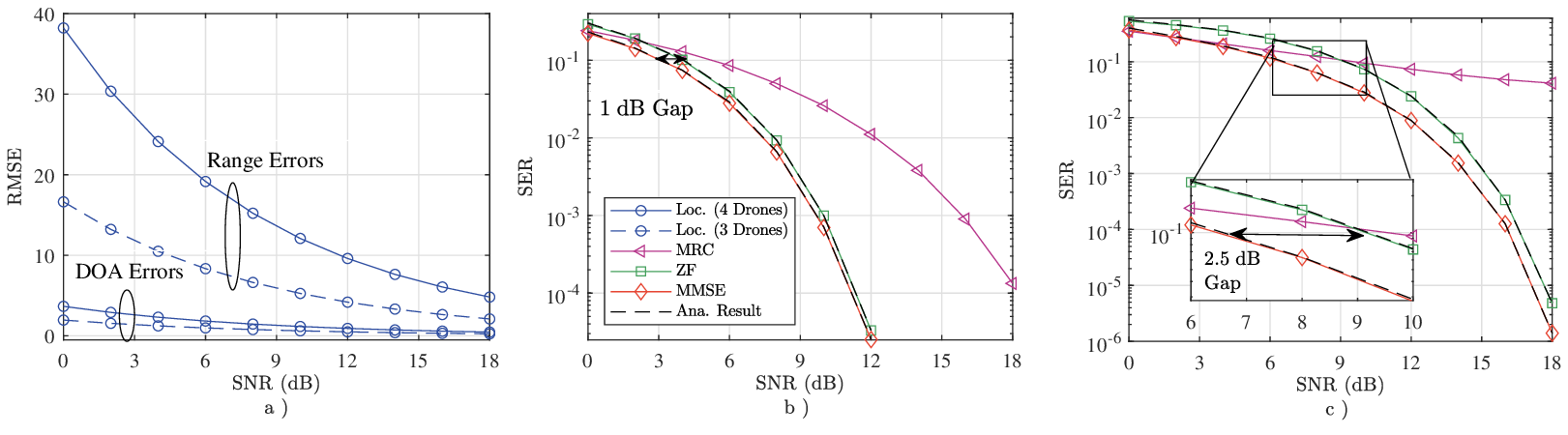}}
\vspace{-0.3cm}
\caption{a) The localization performance with a various number of drones; b) the communication performance with 3 drones; c) the communication performance with 4 drones  } 
\label{SER_gap}
\end{figure*}

  \begin{figure*}[!b]
\hrulefill
\begin{align}
\label{E_9}
\setcounter{equation}{44}
\!\!\!\!\!\!{\mathbb{E}}_9 
&\!\!=\!\! \int_{\Delta f_{D,\min}}^{\Delta f_{D,\max}}
    \underbrace{ 
      \int_{-\pi}^{\pi} 
        \underbrace{ 
          \int_{-\pi}^{\pi} 
            f_{b_3}\!\left( 
              C_1 + C_2 \Delta f_{D,k} 
                 + C_3 \Delta \theta_{k} 
                 + C_4 \Delta \theta_{q_g}
            \right) 
            f(\Delta \theta_k) 
           d(\Delta \theta_k)
        }_{I_1}
        f(\Delta \theta_{q_g}) 
       d(\Delta \theta_{q_g})
    }_{I_2} 
    f(\Delta f_{D,k}) 
   d(\Delta f_{D,k}),
\end{align}
 \vspace{-0.8cm}
\end{figure*}
\begin{figure*}[!b]
 \vspace{-0.3cm}
\begin{align}
\label{E_19}
\setcounter{equation}{62}
    \mathbb{E}_{19}=  
      \int_{-\pi}^{\pi} 
        \underbrace{ 
          \int_{-\pi}^{\pi}f_{b_5}\big(C_{21}  +C_{22}\cos \theta_k\Delta \theta_k
+C_{23}\cos \theta_{\tilde q}\Delta \theta_{\tilde q}\big)f(\Delta \theta_k) 
          \, d(\Delta \theta_k)
        }_{I_{11}} \,
        f(\Delta \theta_{\tilde  q}) 
      \, d(\Delta \theta_{\tilde q}) \,
\end{align}
\end{figure*}

Afterwards, ${{\mathbb{E}}_{15}}$ can be simplified to 
\begin{align}
\setcounter{equation}{54} 
 \mathbb{E}_{15}&\!=\!\frac{1}{2}\sum\limits_{ L = 1}^2\sum\limits_{ r_2 = 1}^2 (-1)^{(L+1)(r_2-1)} j^{L-1}\mathbb{E}_{18},   \vspace{-0.2cm}
 \end{align}
where $\mathbb{E}_{18}$ is given by
 \begin{align}
 \label{E_18}
 \!\!\! \!\!\!\mathbb{E}_{18}\!\!= \!\!\mathbb{E}\bigg[\!f_{L}\!\big(\!C_{15,r_2}  \!\!+\!\!C_{18,r_2}\!\cos \theta_k\Delta \theta_k
\!+\!C_{17,r_2}\!\cos \theta_{q_c}\Delta \theta_{q_c}\!\big)\!\bigg],\!\!\!\!
 \end{align}
 where $f_1=\cos(\cdot)$ and $f_2=\sin(\cdot)$. $C_{15,r_2}=C_{18,r_2}\sin \theta_k+C_{17,r_2}\sin \theta_{q_c} $, in which 
 \begin{subequations}
     \begin{align}
       C_{18,r_2}&=C_{5}\bigg[(i_3-n_2)+(-1)^{r_2+1}(i_2-i_1)\bigg],  \\
       C_{17,r_2}&=C_{5}\bigg[(n_2-i_3)+(-1)^{r_2+1}\sum\limits_{ c = 1}^2 (-1)^c(n_1-i_c)\bigg]. 
     \end{align}
 \end{subequations}

   \begin{figure}[ptb]
\centering
{\includegraphics  [height=3.22in, width=2.97in]{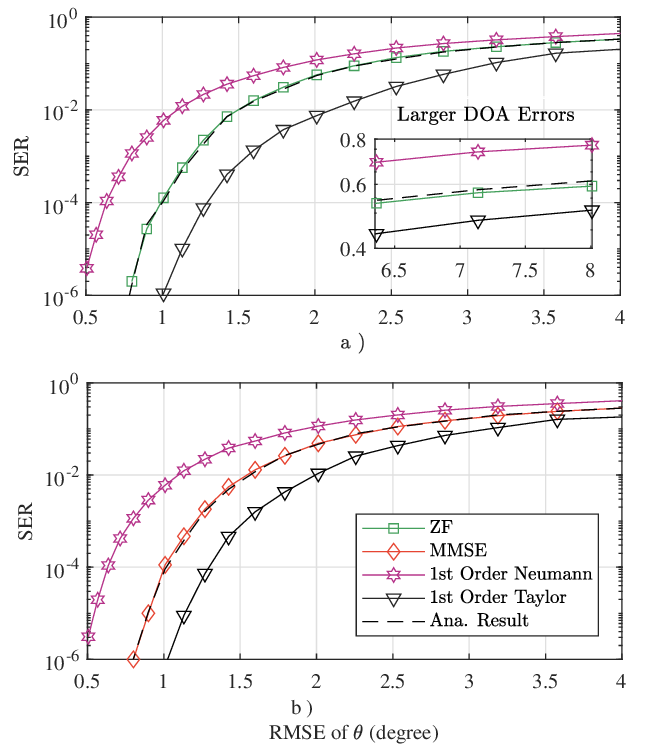}}
\vspace{-0.3cm}
\caption{The approximation performance of the proposed method}
\color{black}
\label{approximation performance}%
\vspace{-0.3cm}
\end{figure}

To obtain ${{\mathbb{E}}_{11}},...,{{\mathbb{E}}_{15}}$, $ \mathbb{E}_{16}$ in \eqref{E_16}, $\mathbb{E}_{13}$ in \eqref{E_13}, $\mathbb{E}_{17}$ in \eqref{E_17} and $\mathbb{E}_{18}$ in \eqref{E_18} should be calculated. The derivation of the closed-form expressions of $\mathbb{E}_{13}$, $\mathbb{E}_{16}$, $\mathbb{E}_{17}$ and $\mathbb{E}_{18}$ can be found in Appendix \ref{Appendix_A}.

In the end, the closed-form expression of ${\mathbb{E}}_{\ell+6}$ in \eqref{E_7} is utilised to obtain ${{\mathbb{E}}[({ \tilde \nu})^{{k_1}}]}$ and ${{\mathbb{E}}[({  \nu_{y}})^{{k_1}}]}$ when $\ell=1$ and $\ell=2$, respectively. The closed-form expression of ${{\mathbb{E}}_{11}},...,{{\mathbb{E}}_{15}}$ can be employed to obtain ${{\mathbb{E}}[{{{\Gamma }}}_{\mathcal{D}}]}$ and ${{\mathbb{E}}[{{{\Gamma }}}^2_{\mathcal{D}}]}$, which are substituted into \eqref{E_sqrt_gammax_k1} to calculate ${{\mathbb{E}}[\big({\sqrt{{{{{\Gamma }}}_{\mathcal{D}}}}}\big)^{{k_1}}]}$. Since ${\mathbb{E}}_5\approx \displaystyle  { {{\mathbb{E}}[({ \tilde \nu})^{{k_1}}]}/{{\mathbb{E}}[\big({\sqrt{{{{{\Gamma }}}_{\mathcal{D}}}}}\big)^{{k_1}}]}}  $ and ${\mathbb{E}}_6\approx \displaystyle \! { {{\mathbb{E}}[({  \nu_{y}})^{{k_1}}]}/{{\mathbb{E}}[\big({\sqrt{{{{{\Gamma }}}_{\mathcal{D}}}}}\big)^{{k_1}}]}}$, the approximated value of ${\mathbb{E}}_5$ and ${\mathbb{E}}_6$ can be obtained, which are then substituted into \eqref{E 3 E 4} to calculate ${{\mathbb{E}}_{3}}$ and ${{\mathbb{E}}_{4}}$. Afterwards, ${{\mathbb{E}}_{3}}$ and ${{\mathbb{E}}_{4}}$ are employed to calculate ${\mathbb{E}}_{1}$ and ${\mathbb{E}}_{2}$ in \eqref{E_ell}. Finally, the average SER can be derived through ${P_{{e}}}  \!\!=\!\!\sum\limits_{{m_1},...,{m_K} \in {S}}\!\! ({  {\mathbb{E}}_1+{\mathbb{E}}_2}
)/{M^{K}}$.


\section{\color{blue}Numerical Results\color{black}}
To validate the analytical derivations and evaluate the PASCAL system's performance with ZF and MMSE, this section provides a comprehensive set of simulation results. The simulation parameters can be found in Table \ref{sim par}, where ${N_\mathrm{MC}}$ refers to the number of Monte-Carlo trials. Each simulation point is obtained by averaging over ${N_\mathrm{MC}}$ tests. Param. indicates parameters. In the following results, the localization performance is evaluated in terms of root mean square error (RMSE), while the communication performance is assessed based on SER. Without loss of generality, the same transmit power is evenly allocated to drones. In Fig. \ref{RMSE_SER} and Fig. \ref{ISAC_PASCAL}, the RMSE value of each location parameter is obtained by averaging the RMSE value corresponding to different drones. Similarly, the SER value is calculated by averaging the SER value corresponding to different drones.

In Fig. \ref{RMSE_SER}, the effect of localization inaccuracies on various equalizers including ZF, MMSE and MRC in the PASCAL system is demonstrated. The signals from drone 1 and drone 2 are received by a BS composed of $N=5$ antennas at $\mathrm{SNR}=12 \; \mathrm{dB}$ and the employed modulation type is QPSK. The magnitude of localization inaccuracies are controlled by varying the number of pilots. In specific, the number of pilots increases from 22 to 40. In Fig. \ref{RMSE_SER}, loc. and ana. are the abbreviations of localization and analytical, respectively. As can be observed from Fig. \ref{RMSE_SER} that localization inaccuracies have a negative effect on SERs of all equalizers since their SERs increase with RMSE of localization errors. By comparing with the ideal localization case, it can be also found that the discrepancy between the actual SER and the ideal SER progressively increases under the effect of larger localization errors. By observing the performance of MRC, it is evident that MRC suffers from the error floor issue and thus its performance almost remains invariant with respect to increased localization errors.
By comparing the performance of ZF and MMSE, the performance of MMSE is superior to that of ZF across the whole range. Furthermore, the simulated and analytical results match perfectly by using 3rd order Neumann approximation and 8th order Taylor approximation.    

In Fig. \ref{other_drones_effect}, the influence of localization inaccuracy of other drones  $\Delta {{ \boldsymbol{\psi}  }}_i$ (i.e., ${\Delta \theta _i}$, $\Delta d_i$ and ${\Delta f_{D,i}} $) for $i \ne k$ on the SER of drone $k$ under various equalizers is demonstrated. A BS composed of $ N=4$ antennas is utilized to serve drone 1 and drone 2 with 30 pilots. Two modulation types including BPSK and QPSK are considered. In Fig. \ref{other_drones_effect}, the SER value of drone 1 is calculated with various SNR, whereas drone 2 is considered as interference ($k=1$, $i=2$). The actual SER value of drone 1 (i.e., $\Delta {{ \boldsymbol{\psi}  }}_2\ne0$) with various equalizers is compared to the ideal case, in which the localization errors of drone 2 equal to zero (i.e., $\Delta {{ \boldsymbol{\psi}  }}_2=0$). If the actual and ideal SER of one equalizer are inconsistent, this indicates that the SER of the equalizer is influenced by the localization errors of drone 2. As can be noted from Fig. \ref{other_drones_effect}, both ZF and MMSE are influenced by the localization inaccuracy of drone 2, whereas MRC is not affected by the errors from drone 2, under both BPSK and QPSK scenarios. For example, the SER of ZF and MMSE are $4.4\times 10^{-4}$ and $3\times 10^{-4}$ at $\mathrm{SNR}=5 \; \mathrm{dB}$ when BPSK is employed, in the ideal case, while they become $1\times 10^{-3}$ and $5\times 10^{-4}$ in the presence of $\Delta  {\boldsymbol{\psi}}_2$. However, the SER of MRC remain $2.15\times 10^{-3}$ both under ideal conditions and in the presence of $\Delta  {\boldsymbol{\psi}}_2$. The simulated and analytical results match well with 3rd order Neumann approximation and 10th order Taylor approximation. 

Fig. \ref{dominant_parameter} verifies the dominant role of DOA estimation errors (i.e., $\Delta \theta$ ) among various localization errors. In Fig. \ref{dominant_parameter}, drone 1 and drone 2 are employed to send signals to a BS consisting of $N=4$ antennas in case I and a BS composed of $N=5$ antennas in case II, respectively. The modulation technique is QPSK and 20 pilots are utilized. Act. is the abbreviations of actual. In Fig. \ref{dominant_parameter}, even though actual SER and ideal SER without localization errors exhibit a substantial disparity, when the DOA estimation errors (i.e., $\Delta \theta$) equal to zero, the resulting SER approaches that of ideal SER. For example, when $\mathrm{SNR}=1 \; \mathrm{dB}$ and $N=4$, the SER of ZF under the scenario with full localization errors, the scenario with ideal $\Delta \theta$ and the scenario with ideal localization is 0.286, 0.148, 0.130. The phenomenon is more obvious when $N=5$. This indicates that the estimation errors for Doppler frequency and range may be ignored in some scenarios, which can simplify algorithmic and system design in these applications. The simulated and analytical results exhibit a close agreement with 3rd order Neumann approximation and 4th order Taylor approximation.

Fig. \ref{ISAC_PASCAL} shows the performance comparison between the PASCAL system and ISAC system in \cite{ISAC_Dong,ISAC_ouyang2}. In ISAC system, a BS is employed to send signals towards the targets and then rely on the reflected signals to complete the localization of targets. In the meanwhile, the BS receives the uplink signals from the users to complete the communication with the users. Given that one of the major difference between PASCAL and ISAC is that communication and localization share the same resources (i.e., transmit power, pilots) in PASCAL, while in the conventional ISAC system, they compete for the same resource, we explore their performance gap given the same amount of resources. To ensure a fair comparison, we assume the targets and users in the ISAC system have the same location since targets and users are identical (i.e., drones are both targets and users) in the PASCAL system. In Fig. \ref{ISAC_PASCAL}, a two-drone case (i.e., drone 1 and drone 2) is considered. A BS composed of $N=4$ antennas is employed at $\mathrm{SNR}=9 \; \mathrm{dB}$ and the modulation type is BPSK. The number of pilots is 100. These pilots and transmit power will be allocated to communication and localization based on predefined percentage ratios. As can be observed from Fig. \ref{ISAC_PASCAL} that the localization performance of ISAC degrades with the increase of comm-loc resource ratio. However, the communciation performance of ISAC gradually increases with the increase of this ratio. This result clearly indicates that the competitive relationship between communication and localization functions for the ISAC system.  Compared to the ISAC system, both localization and commuication performance of the PASCAL system does not change with various comm-loc resource ratio. In addition, its performance is superior to that of the ISAC system throughout the whole range. This indicates that PASCAL does not depedent on the comm-loc resource ratio. In addition, the localization in the conventional ISAC system \cite{ISAC_Dong,ISAC_ouyang2} suffers from a two-way path loss so that it performance may never reach that of PASCAL since PASCAL only has one-way path loss. The simulated and analytical results show excellent agreement when 3rd order Neumann approximation and 8th order Taylor approximation are employed.

Fig. \ref{SER_gap} show the PASCAL system performance with more drones are considered. In Case I, drone 1–3 are employed to transmit signals to a BS with $N=5$ antennas, whereas drone 1–4 are utilized in Case II to transmit signals to the same BS. The employed modulation type is QPSK and the number of pilots is 41. In Fig. \ref{SER_gap}, the RMSE and SER of drone 1 are employed as examples to indicate the communication and localization performance, respectively. As shown in Fig. \ref{SER_gap} a), where estimation errors for DOA and range are employed as examples to characterize the localization performance, localization becomes worse when more drones are considered. This is due to the increased user interference in Case II compared to that in Case I. By comparing Fig. \ref{SER_gap} b) and Fig. \ref{SER_gap} c), it can be found that MRC exhibits a error floor issue in Fig. \ref{SER_gap} c), indicating that the limitation of MRC in cases with a large number of drones. Compared to MRC, the SER of ZF and MMSE gradually reduce throught the whole range, which is not only attributed to the improved SNR condition, but also due to the reduced localization errors. In addition, the performance of ZF is worse that than MRC when the SNR is relatively low. In Fig. \ref{SER_gap}, it can also be noted that the performance gap between ZF and MMSE becomes larger when more drones are considered. This indicates that MMSE has a greater advantage compared with MRC and ZF in cases with more drones and also demonstrates that MMSE exhibits stronger robustness when confronted with the challenges posed by user interference and localization inaccuracy. The simulated and analytical results match well by using 2rd order Neumann approximation and 10th order Taylor approximation.

Fig. \ref{approximation performance} demonstrates the performance of the proposed approximation method combining Neumann approximation and Taylor approximation in Sec. \ref{Approximation Methods}. In Fig. \ref{approximation performance}, the signals from drone 1-3 are received by a BS composed of $N=5$ antennas with a SNR from $-4 \ \mathrm{dB}$ to $12 \ \mathrm{dB}$. The employed modulation type is QPSK and the pilot number is 39. Here, the RMSE and SER of drone 2 are employed as examples to indicate the communication and localization performance, respectively. In Fig. \ref{approximation performance}, our approximation method is compared to the 1st order Neumann approximation method employed in \cite{} with perfect Taylor approximation and the 1st order Taylor approximation method employed in \cite{} with perfect Neumann approximation. As can be observed from Fig. \ref{approximation performance} that our approximated result match the simulated values very well for both ZF and MMSE by using 5rd order Neumann approximation and 11th order Taylor approximation. Nevertheless, the approximate values obtained by using the methods in both \cite{} and \cite{} exhibit a noticeable deviation deviation from the simulated values. The result indicates the superiority of our approximation method. 

\section{Conclusion}
A SER analysis for the PASCAL system with ZF and MMSE equalizers was provided. In specific, multiple drones actively transmitted signals to a BS, which was responsible for extracting the location parameters with the assistance of pilots and conducting the data decoding by incorporating the estimated location parameters into various equalizers. A tightly approximated SER analysis was provided by employing a hybrid method composed of a Neumann approximation and Taylor approximations. The mechanisms through which localization errors affected communication performance with ZF and MMSE was also investigated and was compared to that of MRC. It was found that MRC was unaffected by the localization errors of all drones except for drone $k$, as well as by the range estimation errors, whereas ZF and MMSE were influenced by these errors. In addition, among all error sources, ZF and MMSE are most sensitive to the estimation errors originating from DOA. These findings can provide insights on system design for wireless systems highly rely on location accuracy, such as location-aware service. Numerical simulation results verified the accuracy of our analysis.

\begin{appendices}
\section{ The Complete Derivations of $\mathbb{E}_{9}$, $\mathbb{E}_{13}$ and $\mathbb{E}_{16},...,\mathbb{E}_{18}$}
Given that different variables (i.e., $\Delta {\theta _k}$, $\Delta \theta_{q_g}$ and $\Delta {f_{D,k}} $) are independent with each other \cite{ISAC2_Han}, $\mathbb{E}_{9}$ can also be given in \eqref{E_9} in the bottom of page \pageref{E_9}. In  \eqref{E_9}, ${\mathbb{E}}_9$ can be evaluated by calculating the innermost integral first (i.e., $I_1$), while holding other variables constant. Thereafter, the result of the innermost integral will be integrated with respect to the other two variables ($\Delta \theta_{q_g}$ and $\Delta {f_{D,k}} $). By substituting the expression of $f(\Delta {\theta_k})$ in \cite[eq. (9)]{ISAC2_Han} into \eqref{E_9}, ${I}_{1}$ can be expressed as 
\begin{align}
\setcounter{equation}{57} 
    {I}_{1}&=\displaystyle
  {\frac{1}{\sqrt {2\pi } \sigma_{\theta_k} }}\! \int_{-\pi}^{\pi } \! f_{b_3}(C_{19}\!+\!C_3\Delta \theta_{k}) {{\rm{e}}^{ - \frac{1}{2}{{\left(\frac{{\Delta {\theta _k}}}{\sigma_{\theta_k} }\right)}^2}}}d\Delta \theta_k.
\end{align}

Then, by using Euler's formula to convert the trigonometric function $f_{b_3}(\cdot)$ as shown in \eqref{fb3} into exponential functions, 
${I}_{1}$ can be simplified to 
\begin{align}
\label{I_1_new}
    {I}_{1}={\frac{1}{2\sqrt {2\pi } \sigma_{\theta_k} j^{b_4} } } \left({\rm{e}}^{jC_{19}}I_3+(-1)^{b_4} {\rm{e}}^{-jC_{19}}I_4\right),
\end{align}
where $C_{19}=C_1+C_2\Delta f_{D,k}+C_4\Delta \theta_{q_g}$. $b_4=0$ if $f_{b_3}(\cdot)=\cos(\cdot)$, while $b_4=1$ if $f_{b_3}(\cdot)=\cos(\cdot)$. $I_3$ and $I_4$ can be denoted by using a general expression $I_5$, which is given in \eqref{I_11} when $C_{24}=C_{3}$ and ${\Delta {\psi}}=\Delta {\theta_{{k}}}$. The derivation of $I_{5}$ can be found in Appendix \ref{closed-form expression}. By using the result of $I_5$, $I_3$ and $I_4$ can be obtained, which are then substituted into \eqref{I_1_new} to obtain the closed-form expression of $I_1$. 

The result of $I_1$ can then be employed to obtain $I_2$. By substituting the expression of $f(\Delta {\theta_{q_g}})$ into \eqref{E_9}, $I_2$ can be denoted by 
\begin{align}
    I_2={\frac{1}{\sqrt {2\pi } \sigma_{\theta_{q_g}} }} \int_{-\pi}^{\pi }  I_1 {{\rm{e}}^{ - \frac{1}{2}{{\left(\frac{{\Delta {\theta_{{q_g}}}}}{\sigma_{\theta_{q_g}} }\right)}^2}}}d\Delta \theta_{q_g}
\end{align}

Then, by conducting some simple algebraic operations, $I_2$ can be simplified to 
\begin{align}
    I_2={\frac{1}{ {4\pi } \sigma_{\theta_{k}} \sigma_{\theta_{q_g}} j^{b_4} }} \left({\rm{e}}^{jC_{20}}I_2I_6+(-1)^{b_4} {\rm{e}}^{-jC_{20}}I_3I_7\right), 
\end{align}
where $C_{20}=C_1+C_2\Delta f_{D,k}$. $I_6$ and $I_7$ can be expressed by using the general expression $I_8$, which is given in \eqref{I_11} when $C_{24}=C_{4}$ and ${\Delta {\psi}}=\Delta {\theta_{{q_g}}}$. Similarly, by using the result of $I_8$, the closed-form expression of $I_2$ can be obtained. 

Finally, the result of $I_2$ can be averaged over $\Delta f_{D,k}$ to obtain ${\mathbb{E}}_9$. By substituting the expression of $f(\Delta {\theta_{q_g}})$ into \eqref{E_9}, and performing some algebraic operations, ${\mathbb{E}}_9$ can be simplified to  
\begin{align}
   {\mathbb{E}}_9
   &={\frac{1}{\sqrt {2\pi } \sigma_{f_{D}} }}\! \int_{\Delta f_{D,\min}}^{\Delta f_{D,\max} } \! I_2 {{\rm{e}}^{ - \frac{1}{2}{{\left(\frac{{\Delta {f_{D,k}}}}{\sigma_{f_{D}} }\right)}^2}}}d\Delta f_{D,k}\nonumber \\
    &={\frac{1}{ {4\sqrt{2}\pi^{3/2} } \sigma_{\theta_{k}} \sigma_{\theta_{q_g}} \sigma_{\theta_{f_D}}j^{b_4} }}\bigg({\rm{e}}^{jC_{1}}I_3I_6I_8 \nonumber \\
    &+(-1)^{b_4} {\rm{e}}^{-jC_{1}}I_4I_7I_9\bigg) , 
\end{align}
where $I_8$ and $I_9$ can be expressed by using a general expression $I_{10}$, which is shown in \eqref{I_11} when $C_{24}=C_{2}$ and ${\Delta {\psi}}=\Delta {f_{D,k}}$. By using the result of $I_{10}$, the closed-form expression of ${\mathbb{E}}_9$ can be derived. 

$\mathbb{E}_{13}$ and $\mathbb{E}_{16},...,\mathbb{E}_{18}$ can be denoted by using $\mathbb{E}_{19}$ in \eqref{E_19} shown in the bottom of page \pageref{E_19}, in which $\theta_{\tilde q} \in \{\theta_{ q1 }, \theta_{ qc }\}$. $f_{b5}(\cdot)$ can be either $\cos(\cdot)$ or $\sin(\cdot)$. Similar to  $\mathbb{E}_{9}$, $\mathbb{E}_{19}$ can be evaluated by first calculating the inner integral (i.e., $I_{11}$), which is then integrated with respect to another variable (i.e., $\Delta \theta_{\tilde q}$). The result of $\mathbb{E}_{19}$ can be given by 
\begin{align}
\setcounter{equation}{63} 
    \mathbb{E}_{19}={\frac{1}{ {4\pi } \sigma_{\theta_{k}} \sigma_{\theta_{\tilde q}} j^{b_4} }} \left({\rm{e}}^{jC_{21}}I_{12}I_{15}+(-1)^{b_4} {\rm{e}}^{-jC_{21}}I_{13}I_{16}\right), 
\end{align}
where $I_{12}$ and $I_{13}$ can be denoted by a general expression $I_{14}$, which is shown in \eqref{I_11} when $C_{24}=C_{22} \cos\theta_k$ and ${\Delta {\psi}}=\Delta \theta_k$. $I_{15}$ and $I_{16}$ can be denoted by the general expression $I_{17}$, which can be obtained in \eqref{I_11} when $C_{24}=C_{23} \cos\theta_{\tilde q}$ and ${\Delta {\psi}}=\Delta \theta_{\tilde q}$. 

\section{ The Complete Derivations of $I_5$, $I_8$, $I_{10}$, $I_{14}$ and $I_{17}$}
\label{closed-form expression}
The general expression of $I_5$, $I_8$, $I_{10}$, $I_{14}$ and $I_{17}$ is given by 
\begin{align}
\label{I_11}
    I_{18}=\int_{{\Delta \psi_{\min}}}^{{\Delta \psi_{\max}}}{{\rm{e}}^{ \pm j{C_{24}{\Delta {\psi}}{\rm{ - }}\frac{{\rm{1}}}{{{\rm{2}}\sigma _\psi ^2}}{\Delta {\psi}}^2}}} d {\Delta {\psi}},
\end{align}
where $C_{24} \in \{C_{3},C_{4}, C_2, C_{22} \cos\theta_k, C_{23} \cos\theta_{\tilde q}  \}$.  ${\Delta \psi_{\min}}$ and ${\Delta \psi_{\max}} $ indicate the minimum and maximum values of ${\Delta {\psi}}$, and $\sigma _\psi$ refer to the variance of ${\Delta {\psi}}$. $ {\Delta \psi } \in \{\Delta \theta_k, \Delta \theta_{q_g}, \Delta f_{D,k}, \Delta \theta_k, \Delta \theta_{\tilde q}\}$.

The closed-form expression of $I_{18}$ can be given by 
\begin{align}
    I_{18}=\frac{{\sqrt {2\pi \sigma _\psi ^2} {{\rm{e}}^{ - \frac{{C^2_{21}}\sigma _\psi ^2}{2}}}\left[ {\rm{erf}}(u_{max})-{\rm{erf}}(u_{min}) \right]}}{2},
\end{align}
where $u_{\max}=\frac{  {\sqrt 2} \Delta \psi_{\max} \mp j{\sqrt 2} C_{24}\sigma^2_\psi }{2{\sigma_\psi} }$ and $ u_{\min}=\frac{  {\sqrt 2 {\Delta \psi_{\min}}} \mp j{\sqrt 2 }{C_{24}}\sigma^2_\psi }{2\sigma_\psi }$.
\label{Appendix_A}

 \end{appendices}

\end{document}